\pdfoutput=1
\documentclass[letterpaper,journal]{IEEEtran}
\usepackage{amsmath,amsfonts,amssymb}
\usepackage{algorithmic}
\usepackage{algorithm}
\usepackage{array}
\usepackage[caption=false,font=footnotesize,labelfont=sf,textfont=sf]{subfig}
\usepackage{textcomp}
\usepackage{stfloats}
\usepackage[hyphens]{url}
\usepackage{verbatim}
\usepackage{graphicx}
\usepackage{cite}
\usepackage{booktabs}
\usepackage{bm}
\usepackage{tikz}
\usepackage{pgfplots}
\usetikzlibrary{positioning,arrows.meta,fit,backgrounds,calc}
\usepgfplotslibrary{groupplots,fillbetween}
\pgfplotsset{compat=1.18}
\usepackage[hidelinks]{hyperref}
\usepackage{orcidlink}
\newcommand{\EsMassWeek}{less than 0.01}

\newcommand{\IsoAblArfimadMAPE}{18.5}

\newcommand{\IsoAblBlinddMAPE}{7.8}

\newcommand{\IsoAblBrkdMAPE}{3.1}

\newcommand{\IsoAblFracdMAPE}{3.1}

\newcommand{\IsoAblGbmMAPE}{2.57}

\newcommand{\IsoAblGbmdMAPE}{67.9}

\newcommand{\IsoAblGbmesMAPE}{2.68}

\newcommand{\IsoAblMemdMAPE}{37.0}

\newcommand{\IsoAblOudMAPE}{1.7}

\newcommand{\IsoAblResdMAPE}{54.6}

\newcommand{\IsoAblSingledMAPE}{18.6}

\newcommand{\IsoAlphaC}{0.75}

\newcommand{\IsoAlphaH}{0.57}
\newcommand{\IsoAlphaHmax}{0.62}
\newcommand{\IsoAlphaHmin}{0.53}

\newcommand{\IsoAwareBreakB}{14~September~2008}

\newcommand{\IsoAwareCIB}{3 to 25~September~2008}

\newcommand{\IsoAwareD}{0.10}
\newcommand{\IsoAwareDresid}{0.12}
\newcommand{\IsoAwareM}{4}
\newcommand{\IsoAwarePhi}{0.50}
\newcommand{\IsoBPymax}{14.16}
\newcommand{\IsoBPymin}{11.03}

\newcommand{\IsoBestMAPEName}{LSTM}

\newcommand{\IsoBlindDresid}{0.02}
\newcommand{\IsoBlindM}{6}

\newcommand{\IsoCovEmpDiag}{47.6}
\newcommand{\IsoCovEmpJoint}{6.1}
\newcommand{\IsoCovRatio}{3.36}
\newcommand{\IsoCovViol}{56}

\newcommand{\IsoDMsig}{10}
\newcommand{\IsoDMtot}{10}

\newcommand{\IsoDecSaveBase}{69.8}
\newcommand{\IsoDecSaveDiag}{73.2}

\newcommand{\IsoDecViolE}{4.1}
\newcommand{\IsoDecViolEDiag}{36.7}
\newcommand{\IsoDecViolEModel}{3.2}
\newcommand{\IsoDecViolEModelDiag}{4.6}
\newcommand{\IsoDecViolP}{6.5}
\newcommand{\IsoDnoBreaks}{0.25}

\newcommand{\IsoGainMAPE}{15.7}
\newcommand{\IsoGainPin}{6.8}

\newcommand{\IsoHMassDay}{74}
\newcommand{\IsoHMassWeek}{8.6}

\newcommand{\IsoHSdLog}{2.60}
\newcommand{\IsoHSpread}{13}

\newcommand{\IsoKernelSSERed}{5.3}
\newcommand{\IsoMeanLoad}{14.6}
\newcommand{\IsoMinPerYear}{3.9}
\newcommand{\IsoNalarms}{11}
\newcommand{\IsoNalarmsEvent}{8}
\newcommand{\IsoNdays}{3288}
\newcommand{\IsoNhours}{124{,}200}
\newcommand{\IsoNyears}{9}
\newcommand{\IsoPeakLoad}{27.6}
\newcommand{\IsoPeriod}{Mar.~2003 to Apr.~2017}
\newcommand{\IsoPropMAPE}{1.56}
\newcommand{\IsoPropPICP}{91.9}

\newcommand{\IsoTauH}{7.8}

\newcommand{\PacfMax}{0.0196}

\newcommand{\PjmAblArfimadMAPE}{5.8}

\newcommand{\PjmAblBlinddMAPE}{9.1}

\newcommand{\PjmAblBrkdMAPE}{1.7}

\newcommand{\PjmAblFracdMAPE}{1.2}

\newcommand{\PjmAblGbmMAPE}{2.65}

\newcommand{\PjmAblGbmdMAPE}{24.3}

\newcommand{\PjmAblGbmesMAPE}{2.80}

\newcommand{\PjmAblMemdMAPE}{28.6}

\newcommand{\PjmAblOudMAPE}{7.1}

\newcommand{\PjmAblResdMAPE}{64.1}

\newcommand{\PjmAblSingledMAPE}{5.9}

\newcommand{\PjmAlarmBDate}{5~April~2020}
\newcommand{\PjmAlarmBDelta}{-3.89}

\newcommand{\PjmAlarmBStart}{9~March~2020}
\newcommand{\PjmAlarmCDate}{22~June~2020}
\newcommand{\PjmAlarmCDelta}{4.42}

\newcommand{\PjmAlarmCStart}{2~June~2020}

\newcommand{\PjmAlphaC}{0.85}

\newcommand{\PjmAlphaH}{0.45}
\newcommand{\PjmAlphaHmax}{0.54}
\newcommand{\PjmAlphaHmin}{0.50}

\newcommand{\PjmAwareD}{0.20}

\newcommand{\PjmAwareM}{1}

\newcommand{\PjmBPymax}{79.95}
\newcommand{\PjmBPymin}{63.99}

\newcommand{\PjmBestMAPEName}{GBDT with load lags}

\newcommand{\PjmBlindM}{6}

\newcommand{\PjmCovEmpDiag}{55.6}
\newcommand{\PjmCovEmpJoint}{5.2}
\newcommand{\PjmCovRatio}{3.85}
\newcommand{\PjmCovViol}{61}

\newcommand{\PjmDMsig}{10}
\newcommand{\PjmDMtot}{10}

\newcommand{\PjmDecSaveBase}{68.7}
\newcommand{\PjmDecSaveDiag}{66.9}

\newcommand{\PjmDecViolE}{8.5}
\newcommand{\PjmDecViolEDiag}{57.5}
\newcommand{\PjmDecViolEModel}{5.3}
\newcommand{\PjmDecViolEModelDiag}{7.6}
\newcommand{\PjmDecViolP}{5.1}

\newcommand{\PjmGainMAPE}{10.3}
\newcommand{\PjmGainPin}{9.1}

\newcommand{\PjmKernelSSERed}{6.4}
\newcommand{\PjmMeanLoad}{89.2}
\newcommand{\PjmMinPerYear}{3.1}

\newcommand{\PjmNdays}{1065}
\newcommand{\PjmNhours}{43{,}080}

\newcommand{\PjmPeakLoad}{151.6}
\newcommand{\PjmPeriod}{Jan.~2017 to Nov.~2021}
\newcommand{\PjmPropMAPE}{2.13}
\newcommand{\PjmPropPICP}{91.6}

\newcommand{\PjmTauH}{4.8}

\newcommand{\PriceRatio}{0.444}
\newcommand{\SimArlAwareMin}{1813}
\newcommand{\SimArlBlindAtThree}{75}

\newcommand{\SimAwareBreaks}{0.1}
\newcommand{\SimBlindBreaks}{3.3}
\newcommand{\SimDelayExcessMax}{55}
\newcommand{\SimDelayExcessMin}{5}
\newcommand{\SimDjointMax}{0.07}
\newcommand{\SimDnaiveMax}{0.38}

\newcommand{\SimSlopeThree}{2.76}
\newcommand{\SimSlopeTwo}{1.46}
\newcommand{\SimSlopeZero}{1.05}
\newcommand{\ZoneAmtlDA}{4.95}

\newcommand{\ZoneAmtlRep}{4.61}

\newcommand{\ZoneProp}{2.30}
\newcommand{\ZoneWins}{8}

\newtheorem{theorem}{Theorem}
\newtheorem{proposition}{Proposition}
\newtheorem{lemma}{Lemma}
\newtheorem{assumption}{Assumption}
\newtheorem{example}{Example}

\newcommand{\HD}{\mathrm{HD}}
\newcommand{\CD}{\mathrm{CD}}
\newcommand{\E}{\mathbb{E}}
\newcommand{\Prob}{\mathbb{P}}
\newcommand{\Var}{\operatorname{Var}}
\newcommand{\Cov}{\operatorname{Cov}}
\newcommand{\T}{\top}
\newcommand{\one}{\mathbf{1}}
\newcommand{\Vind}{V^{\mathrm{ind}}}


\pgfplotsset{
  every axis/.append style={font=\scriptsize, line width=0.5pt, tick style={line width=0.4pt},
    every axis label/.append style={font=\scriptsize}, legend style={font=\tiny, draw=none, fill=none},
    legend cell align=left},
  isoH/.style={color=blue!75!black, thick},
  isoC/.style={color=red!75!black, thick},
  pjmH/.style={color=blue!75!black, thick, densely dashed},
  pjmC/.style={color=red!75!black, thick, densely dashed},
}
\tikzset{
  ciband/.style={fill=blue!12},
  blindbreak/.style={red!70!black, densely dashed, line width=0.6pt},
  awarebreak/.style={blue!70!black, line width=0.6pt},
}

\hypersetup{hypertexnames=false}
\usepackage{multicol}
\makeatletter
\newcommand{\supplementlayout}{%
  \clearpage\onecolumn
  \def\normalsize{\@setfontsize{\normalsize}{11}{13.3846pt}}%
  \def\small{\@setfontsize{\small}{10}{12pt}}%
  \def\footnotesize{\@setfontsize{\footnotesize}{9}{10.5pt}}%
  \def\scriptsize{\@setfontsize{\scriptsize}{8}{9pt}}%
  \def\tiny{\@setfontsize{\tiny}{6}{7pt}}%
  \def\sublargesize{\@setfontsize{\sublargesize}{12}{14pt}}%
  \def\LARGE{\@setfontsize{\LARGE}{17}{20pt}}%
  \def\huge{\@setfontsize{\huge}{20}{24pt}}%
  \def\Huge{\@setfontsize{\Huge}{24}{28pt}}%
  \@IEEEnormalsizeunitybaselineskip=13.3846pt
  \CLASSINFOnormalsizebaselineskip=13.3846pt
  \CLASSINFOnormalsizeunitybaselineskip=13.3846pt
  \normalsize
  \abovedisplayskip=7.42499pt plus 5pt minus 3pt
  \belowdisplayskip=\abovedisplayskip
  \abovedisplayshortskip=0pt plus 5pt
  \belowdisplayshortskip=7.42499pt plus 5pt minus 3pt
  \jot=3.34615pt \IEEEnormaljot=3.34615pt
  \textheight=695.99908pt \topmargin=-52.59776pt \headsep=20.29163pt \footskip=26.50026pt \topskip=13.3846pt
  \global\@colht=\textheight \global\@colroom=\textheight \global\vsize=\textheight
  \parindent=11pt \parskip=0pt plus 0.98997pt minus 0.49503pt
  \abovecaptionskip=6.69229pt
  \textfloatsep=20.74615pt plus 2.67686pt minus 5.35374pt
  \floatsep=11.37698pt plus 2.67686pt minus 2.67686pt
  \intextsep=\floatsep \dbltextfloatsep=\textfloatsep \dblfloatsep=\floatsep
  \@fpsep=10.03844pt plus 2fil \@dblfpsep=\@fpsep
  \footnotesep=8.40002pt \skip\footins=12.04605pt plus 5.35374pt minus 2.67686pt
  \labelsep=5.5pt \leftmargini=22pt \leftmarginii=11pt \leftmarginiii=16.5pt \leftmargin=\leftmargini
  \topsep=6.69229pt \itemindent=-11pt \maxdepth=6.69229pt \@maxdepth=\maxdepth
  \heavyrulewidth=0.88002pt \lightrulewidth=0.55003pt \cmidrulewidth=0.32999pt
  \belowrulesep=3.21745pt \aboverulesep=1.97997pt \cmidrulekern=5.5pt \defaultaddspace=5.5pt
  \fussy\raggedbottom}
\newcommand{\papersizes}{%
  \def\normalsize{\@setfontsize{\normalsize}{10}{12.00pt}}%
  \def\small{\@setfontsize{\small}{9}{10pt}}%
  \def\footnotesize{\@setfontsize{\footnotesize}{8}{9pt}}%
  \def\scriptsize{\@setfontsize{\scriptsize}{7}{8pt}}%
  \def\tiny{\@setfontsize{\tiny}{5}{6pt}}%
  \normalsize
  \parindent=10pt \parskip=0pt plus 0.89998pt minus 0.45001pt
  \labelsep=5pt \leftmargini=20pt \leftmargin=\leftmargini \topsep=6pt \itemindent=-10pt}
\makeatother

\begin{document}

\title{Physics-Informed Fractional Load Forecasting With Memory-Aware Break Detection for Reserve Scheduling}

\author{Ali~Nikseresht\,\orcidlink{0000-0002-6107-7699},~\IEEEmembership{Member,~IEEE}%
\thanks{Ali Nikseresht is with the School of Computer Science, Georgia Institute of Technology, Atlanta, GA 30332 USA, and also with the Rotterdam School of Management, Erasmus University, 3062 PA Rotterdam, The Netherlands (e-mail: ali.nikseresht@ieee.org).}}

\markboth{}%
{Nikseresht: Physics-Informed Fractional Load Forecasting With Memory-Aware Break Detection}

\maketitle

\begin{abstract}
Day-ahead load forecasts set balancing reserves, including the energy a pumped-storage plant holds against forecast errors. Both depend on three effects that are usually modeled separately, namely the thermal inertia of buildings, which makes load lag temperature, structural breaks, which shift demand, and the long memory of forecast errors, which keeps them from cancelling out. We prove that averaging building responses over Cole--Cole distributed time constants yields a fractional temperature kernel with nonnegative weights exactly when its order is at most one. A Bayesian information criterion that ignores long memory selects spurious breaks with probability tending to one, so breaks and memory are estimated jointly, and an online break detector keeps its false-alarm rate below any prescribed level. Long memory also makes the required storage energy grow faster than the square root of the horizon, so a second-order cone program schedules reserves from the joint distribution of hourly errors. On New England and Mid-Atlantic data including the 2020 pandemic, the proposed framework improves on the best of ten benchmarks by \IsoGainMAPE\% and \PjmGainMAPE\% in mean absolute percentage error. Its reserves breach the storage energy limit on \IsoDecViolE\% and \PjmDecViolE\% of days, against \IsoDecViolEDiag\% and \PjmDecViolEDiag\% with error correlations ignored.
\end{abstract}

\begin{IEEEkeywords}
Load forecasting, fractional calculus, long memory, structural breaks, change detection, probabilistic forecasting, reserve scheduling, pumped storage.
\end{IEEEkeywords}

\section*{Nomenclature}
\begin{IEEEdescription}[\IEEEusemathlabelsep\IEEEsetlabelwidth{$\mathcal R_j,k_j,m,\ell$}]
\item[$t,D,h(t)$] Hour, operating day and hour of day.
\item[$L_t,T_t$] Load (GW) and temperature ($^\circ$C).
\item[$\bm L_D$] Loads of the 24 hours of day $D$.
\item[$\mu_t,r_t$] Mean model and residual (GW).
\item[$u^{\HD}_t,u^{\CD}_t$] Heating and cooling degree-hours ($^\circ$C).
\item[$F^{\HD}_t,F^{\CD}_t$] Fractionally smoothed degree-hours ($^\circ$C).
\item[$w_n,\bar w_n$] Kernel weight and tail mass at lag $n$.
\item[$\alpha,\bar\tau$] Fractional order, median time constant (h).
\item[$\beta^{\HD}_h,\beta^{\CD}_h$] Thermal sensitivities (GW/$^\circ$C).
\item[$\mathcal R_j,k_j,m,\ell$] Regime, break day, breaks, minimum length.
\item[$\vartheta_h,y_t$] Hourly scale and standardized residual.
\item[$\kappa,\sigma,H$] Rate, scale and Hurst exponent (Lemma~\ref{lem:fou}).
\item[$d,\phi$] Memory and autoregressive parameters.
\item[$\hat{\bm\mu}_D,\hat{\bm\Sigma}_D$] Predictive mean and covariance of $\bm L_D$.
\item[$\bar r_D,a_D$] Daily mean residual and its innovation.
\item[$\varsigma^2,\psi_k$] Innovation variance, signature of a shift.
\item[$A,\delta,\nu,N_A$] Alarm level, shift, change day, alarm day.
\item[$e_h,b_h$] Hourly imbalance (MW) and storage share.
\item[$R_h,S_h$] Thermal and storage power reserves (MW).
\item[$E,\bar P,\bar E$] Energy headroom, power and energy limits.
\item[$\pi^R_h,\pi^S,\pi^E$] Prices of $R_h$, $S_h$ and $E$.
\item[$\varepsilon_P,\varepsilon_E$] Risk levels of power and energy limits.
\end{IEEEdescription}

\section{Introduction}\label{sec:intro}
\IEEEPARstart{L}{oad} forecasts for the next operating day determine its energy schedule and balancing reserves. For this purpose, a forecast must follow the delayed response of load to temperature, stay accurate when conditions depart from those seen in training, as during a recession or a pandemic, and describe the joint distribution of the 24 hourly errors correctly, because an energy-limited resource such as a pumped-storage plant absorbs the cumulative error of the day and not only its hourly peaks~\cite{matos2011,dispatch2025}. These requirements reflect three effects, namely the thermal inertia of buildings, structural breaks and the long memory of forecast errors.

Recent load forecasting studies have improved accuracy with attention and decomposition networks~\cite{forwardformer2024,tadnet2024,transbilstm2025}, interpretable Kolmogorov--Arnold networks~\cite{kan2025}, multitask and mixture-of-experts learners~\cite{multitask2024,camoe2026}, diffusion models~\cite{diffload2025} and explainable uncertainty propagation~\cite{explain2026}. Changing conditions are handled by online adaptation, from Kalman-filtered additive models~\cite{obst2021,devilmarest2024} to adaptive single-task and multitask learning~\cite{alvarez2021,zaballa2026} and online-offline kernel methods~\cite{dkl2024}. Forecasts are also increasingly judged by the cost of the decisions they support~\cite{cos2025,dfl2026} and embedded in chance-constrained operation~\cite{ccdcopf2026,cfdsuc2024}. The three effects named above, however, are still modeled in isolation, and the gaps between them motivate this work.

\textit{Thermal inertia.} Forecasters represent the delayed response of load to temperature with lagged temperatures and moving averages~\cite{hong2010,wang2016} or with a single exponential smoothing constant, and physics-informed models embed network or device equations~\cite{pisagt2026,upinn2025}. Fractional dynamics have also been learned with physics-informed neural networks in other sectors~\cite{fracpinn2026}. None of these approaches derives the thermal memory of a heterogeneous building stock, and none restricts the learned lag weights to physically admissible values.

\textit{Long memory and structural breaks.} Load forecast errors are persistent, and their level moves with economic regimes. Fractional models~\cite{cheridito2003} and break-adaptive fractional models of energy demand~\cite{nikseresht2024,nikseresht2025} capture each feature, and change-point detection is a mature field~\cite{bai2003,romano2024,song2024}. The two features are easy to confuse in both directions, since level shifts make short-memory data look persistent~\cite{diebold2001} and persistence makes information criteria and sequential detectors that ignore it report breaks that do not exist (Sections~\ref{sec:breaks} and~\ref{sec:mc}). To our knowledge, no load forecasting method estimates breaks and memory jointly or guarantees a prescribed false-alarm rate when the errors have long memory.

\textit{Error covariance and energy-limited reserves.} Probabilistic load forecasts are scored hour by hour~\cite{hong2019,zaballa2026}, and reserve sizing and storage-based compensation of forecast errors rely on these marginal distributions~\cite{matos2011,netloadcomp2024}. The energy that a storage plant must hold, however, depends on all covariances among the hourly errors~\cite{dispatch2025}, and how memory scales this energy has not been quantified.

\begin{figure}[!t]
\centering
\begin{tikzpicture}[font=\scriptsize, >={Stealth[length=3.2pt]}, node distance=2.4mm and 4.5mm,
  box/.style={draw, rounded corners=1.5pt, align=center, text width=37mm, minimum height=7.2mm, inner sep=1.1pt},
  ann/.style={box, fill=gray!15}, day/.style={box, fill=white}]
\node[box, text width=79.5mm, minimum height=5mm] (data) {Hourly load $L_t$, temperature $T_t$, dew point and calendar};
\node[ann, below=4.6mm of data.south west, anchor=north west] (bb) {\textbf{\ref{sec:backbone}} Backbone with kernel \eqref{eq:fes},\\ exact solution of \eqref{eq:qp}};
\node[ann, right=of bb] (br) {\textbf{\ref{sec:breaks}} Breaks and memory by\\ Algorithm~\ref{alg:joint} on \eqref{eq:daily}};
\node[ann, below=of bb] (mu) {\textbf{\ref{sec:backbone}} Boosted correction, mean $\mu_t$\\ and residuals $r_t=L_t-\mu_t$};
\node[ann, right=of mu] (res) {\textbf{\ref{sec:resid}} Residual model \eqref{eq:xpm}\\ with covariance \eqref{eq:fouacov}};
\node[day, below=5.6mm of mu] (law) {\textbf{\ref{sec:resid}} Day-ahead law \eqref{eq:predictive}\\ with $\hat{\bm\mu}_D$ and $\hat{\bm\Sigma}_D$};
\node[day, right=of law] (det) {\textbf{\ref{sec:detector}} Detector \eqref{eq:glr} corrects\\ the level after a shift};
\node[day, below=of law] (socp) {\textbf{\ref{sec:reserves}} Reserve program \eqref{eq:socp}\\ gives $b_h$, $R_h$, $S_h$ and $E$};
\node[day, right=of socp] (price) {Prices $\pi^R_h$, $\pi^S$, $\pi^E$ and\\ plant limits $\bar P$, $\bar E$};
\begin{scope}[on background layer]
\node[draw, densely dashed, rounded corners=2pt, inner sep=1.3mm, fit=(bb)(br)(mu)(res)] (gA) {};
\node[draw, densely dashed, rounded corners=2pt, inner sep=1.3mm, fit=(law)(det)(socp)(price)] (gD) {};
\end{scope}
\node[anchor=south east, inner sep=0.8pt, font=\tiny\itshape] at (gA.north east) {annual estimation, every 1 January};
\node[anchor=north east, inner sep=0.8pt, font=\tiny\itshape] at (gD.south east) {daily operation, at 10:00 of day $D-1$};
\draw[->] (data.south -| bb.north) -- (bb.north);
\draw[->] ([yshift=1.3mm]bb.east) -- ([yshift=1.3mm]br.west);
\draw[->] ([yshift=-1.3mm]br.west) -- ([yshift=-1.3mm]bb.east);
\draw[->] (bb) -- (mu);
\draw[->] (mu) -- (res);
\draw[->] (mu) -- (law);
\draw[->] (res.south west) -- (law.north east);
\draw[->] (res) -- (det);
\draw[->] (det) -- (law);
\draw[->] (law) -- (socp);
\draw[->] (price) -- (socp);
\end{tikzpicture}
\caption{Structure of the framework. The shaded layers are estimated once a year and the white layers run at each issue time. The break layer returns the regimes and the training window of the backbone. Labels give the subsection that develops each layer, and the numbers in parentheses its equations.}
\label{fig:framework}
\end{figure}
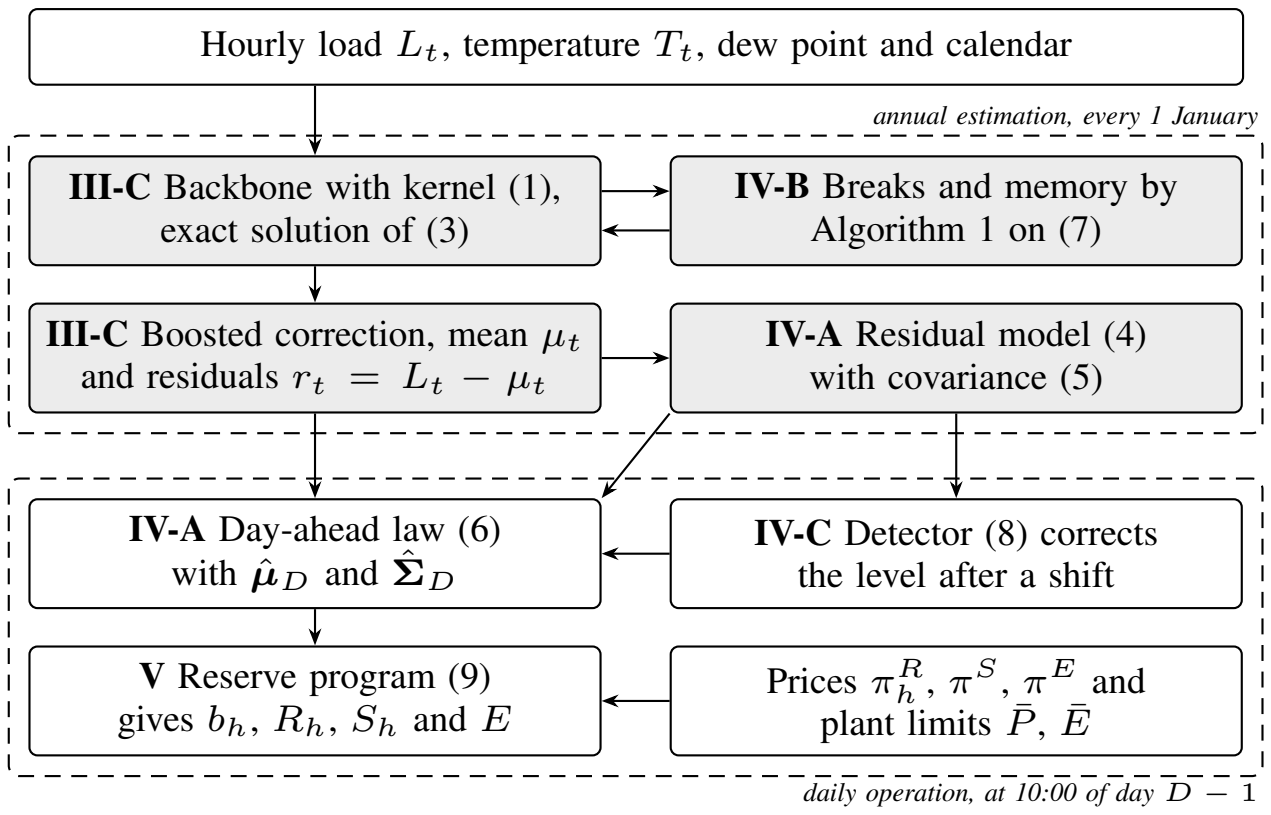

This paper closes the three gaps within one forecasting and scheduling framework, whose layers are outlined in Fig.~\ref{fig:framework}, and makes the following contributions.

\begin{enumerate}
\item \textit{Fractional thermal physics.} We show that aggregating first-order building envelopes whose time constants follow the Cole--Cole law yields a fractional relaxation response. Its discrete form, fractional exponential smoothing, is the average of the exponential smoothing kernels of the individual buildings, has nonnegative weights summing to one exactly when the fractional order satisfies $\alpha\le1$, and decays as a power law. The kernel enters a sign-constrained regression that is solved exactly, and on both systems studied the estimated heating orders are close to one half, far from ordinary exponential smoothing ($\alpha=1$).
\item \textit{Memory-aware structural breaks.} We prove that an information criterion that ignores long memory selects spurious breaks with probability tending to one, and we propose a joint break and memory estimator that provably terminates. Between annual re-estimations, an online mixture likelihood ratio detector built on the innovations of the long-memory residual model and run with threshold $A$ has a mean time to false alarm of at least $A$ days for every memory parameter $d$. It detects a level shift of fixed size after a delay of order $(\log A)^{1/(1-2d)}$ days, which grows with $d$, yet the excess squared one-step error accumulated before the alarm is $2\varsigma^2\log A$, with $\varsigma^2$ the innovation variance, regardless of $d$ and the size of the shift.

\item \textit{Joint predictive laws for energy-limited reserves.} A multiscale fractional Ornstein--Uhlenbeck (fOU) model of the residuals has a closed-form covariance and gives exact Gaussian predictive laws for the 24 hours of the next day. We prove that the storage energy needed to absorb the cumulative error of $n$ hours grows as $n^{H}$, where $H>\frac12$ is the Hurst exponent, so that storage sized from hourly variances fails with probability tending to one, and we schedule reserves with a chance-constrained second-order cone program that uses the joint law.
\item \textit{Evidence on two systems.} The framework is compared with ten benchmarks, from classical regressions to recent neural networks, on fourteen years of ISO New England (ISO-NE) data and five years of PJM Interconnection (PJM) data that include the COVID-19 pandemic. The study includes an ablation of every component, a comparison with adaptive multitask learning~\cite{zaballa2026} on the zones of the Global Energy Forecasting Competition 2017 (GEFCom2017), Monte Carlo checks of the theory, and reserve schedules for two real pumped-storage plants priced with market data.
\end{enumerate}

Section~\ref{sec:setting} states the problem, Sections~\ref{sec:physics} to~\ref{sec:reserves} develop the physics, memory and reserve layers, Section~\ref{sec:cases} reports the case studies and Section~\ref{sec:conclusion} concludes. Complete proofs, numerical checks of every closed form, all settings, and additional results are given in the supplementary material.

\section{Problem Setting}\label{sec:setting}
Let $t$ index hours and $D$ days, and let $h(t)\in\{0,\dots,23\}$ and $D(t)$ be the hour of day and the day of hour $t$. The forecast for operating day $D$ is issued at 10:00 of day $D-1$, when loads up to hour ending 10 are known, and targets the vector $\bm L_D=(L_{24D},\dots,L_{24D+23})$ of the 24 hourly loads of day $D$, that is, lead times from 15 to 38~h. As in GEFCom2017~\cite{hong2019} and in the benchmark studies, the temperatures of day $D$ are taken as observed, so that differences between methods reflect modeling rather than weather forecast quality. All models are re-estimated every 1 January on the data available at that time. The task is to deliver the predictive distribution of $\bm L_D$ and the reserve schedule that it implies.

We write the load as $L_t=\mu_t+r_t$, where the mean $\mu_t$ carries the thermal response, the calendar effects and the current economic regime, and the residual $r_t$ carries the persistent forecast errors, a decomposition that the following three assumptions make precise.

\begin{assumption}[Building envelopes]\label{as:rc}
Building $i$ draws the electric power $q_i\int_0^\infty g_{\tau_i}(s)\,u(t-s)\,ds$ for heating or cooling, where $u$ is the degree-hour forcing, $g_{\tau}(s)=\tau^{-1}e^{-s/\tau}$ is the response of a first-order envelope whose time constant $\tau$ is the product of its thermal resistance and capacitance, and the time constants $\tau_i$ are independent of the gains $q_i$.
\end{assumption}

\begin{assumption}[Regimes]\label{as:regimes}
The level and the thermal sensitivities of $\mu_t$ are piecewise constant in time, with $m$ breaks at days $k_1<\dots<k_m$ and regimes of at least $\ell$ days.
\end{assumption}

\begin{assumption}[Residuals]\label{as:resid}
Within a regime, $r_t=\vartheta_{h(t)}y_t$, where $\vartheta_h>0$ is an hourly scale and $y_t$ is a zero-mean stationary Gaussian process of fOU components (Section~\ref{sec:resid}).
\end{assumption}

Assumption~\ref{as:rc} leads to the thermal kernel of Section~\ref{sec:physics}, Assumption~\ref{as:regimes} to the break and detection layer of Section~\ref{sec:memory} and Assumption~\ref{as:resid} to the predictive law that drives the reserve program of Section~\ref{sec:reserves}, and Fig.~\ref{fig:framework} shows how these layers exchange information.

\section{Fractional Thermal Physics}\label{sec:physics}
\subsection{From Building Envelopes to a Fractional Kernel}
Under Assumption~\ref{as:rc} and for a large building stock, the law of large numbers gives the aggregate demand $\bar q\int_0^\infty g(s)u(t-s)\,ds$, where $\bar q=\sum_iq_i$ and $g(s)=\E[g_{\tau_i}(s)]$ is the average response. A building stock mixes light and heavy construction, so its time constants spread over orders of magnitude. The next result identifies a dispersion law under which the average response is that of a fractional system.

\begin{proposition}[Aggregation]\label{prop:colecole}
Let $V_i=\ln(\tau_i/\bar\tau)$ have the Cole--Cole density $f_\alpha(v)=\sin(\alpha\pi)/\{2\pi[\cosh(\alpha v)+\cos(\alpha\pi)]\}$ with $\alpha\in(0,1)$. Then
(a) $\int_0^\infty e^{-ps}g(s)\,ds=1/[1+(\bar\tau p)^{\alpha}]$ for $p>0$, so the aggregate thermal state $\chi=g\ast u$ solves the fractional relaxation equation $\bar\tau^{\alpha}\,{}^{C}\!D^{\alpha}\chi+\chi=u$, where ${}^{C}\!D^{\alpha}$ is the Caputo derivative and the past is quiescent.
(b) $g(s)=-\frac{d}{ds}E_\alpha\big(-(s/\bar\tau)^{\alpha}\big)$, with $E_\alpha$ the Mittag-Leffler function, is completely monotone, integrates to one and decays as $g(s)\sim\alpha\bar\tau^{\alpha}s^{-1-\alpha}/\Gamma(1-\alpha)$.
(c) $\bar\tau$ is the median time constant, and $\Var(\ln\tau_i)=\pi^2(\alpha^{-2}-1)/3$ vanishes only in the limit $\alpha\to1$ of a homogeneous first-order stock.
\end{proposition}

Part (a) identifies the Stieltjes transform of $f_\alpha$ with $1/(1+x^\alpha)$, and (b) and (c) follow from (a) by Laplace-transform and Tauberian arguments~\cite{feller1971}. The order $\alpha$ thus measures the heterogeneity of the stock and $\bar\tau$ its typical inertia. Averaging many exponential responses produces a scale-free one, the thermal counterpart of Granger's aggregation argument for long memory~\cite{granger1980}, and the dispersion law is the one that Cole and Cole found for dielectric relaxation~\cite{cole1941}.

\subsection{Fractional Exponential Smoothing}
Sampling hourly and replacing the Laplace variable $p$ by the backward difference $1-z$ gives the transfer function
\begin{equation}
W_{\alpha,c}(z)=\frac{1}{1+c(1-z)^{\alpha}}=\sum_{n\ge0}w_nz^n,\qquad c=\bar\tau^{\alpha},\label{eq:fes}
\end{equation}
with $\bar\tau$ in hours, and the filtered forcing $F_t=\sum_{n\ge0}w_nu_{t-n}$. The weights follow from the power series of $(1-z)^\alpha$ by a simple recursion. The weights sum to $W_{\alpha,c}(1)=1$, so $F_t$ is a physically admissible weighted average of past forcing whenever they are nonnegative, and Theorem~\ref{thm:fes} shows that this holds when $\alpha\le1$.

\begin{theorem}[Fractional exponential smoothing]\label{thm:fes}
Let $c>0$, $\alpha\in(0,2)$ and $\bar w_n=\sum_{k>n}w_k$.
(i) If $\alpha<1$, then $w_n=\E[\lambda_V(1-\lambda_V)^n]$, where $\lambda_V=1/(1+\bar\tau e^{V})$ with $\bar\tau=c^{1/\alpha}$ and $V$ has the Cole--Cole density of Proposition~\ref{prop:colecole}. Hence the weights are positive, sum to one and are completely monotone in $n$. If $\alpha=1$, then $w_n=\lambda(1-\lambda)^n$ with $\lambda=1/(1+c)$, which is exponential smoothing.
(ii) If $1<\alpha<2$, then $w_n<0$ for some $n$.
(iii) If $\alpha<1$, then $\bar w_n\sim c\,n^{-\alpha}/\Gamma(1-\alpha)$ and $w_n\sim\alpha c\,n^{-1-\alpha}/\Gamma(1-\alpha)$, so the mean lag is infinite, whereas $\bar w_n=(1-\lambda)^{n+1}$ if $\alpha=1$.
\end{theorem}

Part (i) states that discretization and aggregation commute, since a first-order envelope with time constant $\tau$, discretized by the backward difference, is exponential smoothing with constant $1/(1+\tau)$, and averaging these kernels over the Cole--Cole law gives exactly \eqref{eq:fes}. Part (ii) follows from Abel's theorem~\cite{feller1971}, because nonnegative weights would have a zero mean lag when $\alpha>1$, and it excludes such orders on physical grounds, since a past cold spell would then lower the present heating demand. Part (iii) follows from Karamata's Tauberian theorem~\cite{feller1971}. Proposition~\ref{prop:colecole} produces only orders $\alpha\le1$, and Theorem~\ref{thm:fes} shows that discretization admits no larger order.

\begin{example}[Catalan kernel]\label{ex:catalan}
For $\alpha=\frac12$ and $c=1$, $W(z)=(1-\sqrt{1-z})/z$ and $w_n=C_n/(2\cdot4^n)$, where $C_n$ are the Catalan numbers, so that $w=(\tfrac12,\tfrac18,\tfrac1{16},\tfrac5{128},\dots)$. Half of the weight falls on the current hour, as in exponential smoothing with $\lambda=\tfrac12$, yet 19.6\% of the weight lies beyond a lag of 7~h, against 0.4\% for exponential smoothing, and $w_n\sim n^{-3/2}/(2\sqrt\pi)$, as Theorem~\ref{thm:fes}(iii) predicts.
\end{example}

\subsection{Physics Backbone and Boosted Correction}\label{sec:backbone}
Let $u^{\HD}_t=(T^{\HD}-T_t)^+$ and $u^{\CD}_t=(T_t-T^{\CD})^+$ be the heating and cooling degree-hours with base temperatures $T^{\HD}=15.5\,^\circ$C and $T^{\CD}=18.5\,^\circ$C, and let $F^{\HD}=W_{\alpha_{\HD},c_{\HD}}(B)u^{\HD}$ and $F^{\CD}=W_{\alpha_{\CD},c_{\CD}}(B)u^{\CD}$, where $B$ is the backshift operator. The backbone of the mean model is
\begin{multline}
\mu^{\mathrm b}_t=\bm x_t^\T\bm\omega+\beta^{\HD}_{h(t)}F^{\HD}_t+\beta^{\CD}_{h(t)}F^{\CD}_t\\
+\sum_{j\ge2}\one\{t\in\mathcal R_j\}\big(\Delta\mu_j+\Delta\beta^{\HD}_jF^{\HD}_t+\Delta\beta^{\CD}_jF^{\CD}_t\big),\label{eq:backbone}
\end{multline}
where $\bm x_t$ holds hour-of-week and holiday-by-hour indicators, three annual harmonics for each hour of day and a linear trend, with coefficients $\bm\omega$, and $\mathcal R_1,\mathcal R_2,\dots$ are the regimes in the training window (Section~\ref{sec:breaks}). Each later regime $j\ge2$ receives a level shift $\Delta\mu_j$ and sensitivity changes $\Delta\beta^{\HD}_j$ and $\Delta\beta^{\CD}_j$ relative to the first one. With $\bm\theta=(\alpha_{\HD},\bar\tau_{\HD},\alpha_{\CD},\bar\tau_{\CD})$, the parameters solve
\begin{equation}
\min\ \sum_t\big(L_t-\mu^{\mathrm b}_t\big)^2\quad\text{s.t.}\quad\beta^{\HD}_h\ge0,\ \beta^{\CD}_h\ge0,\ \bm\theta\in\Theta,\label{eq:qp}
\end{equation}
where $\Theta$ is the box in which each order lies in $[0.05,1]$ and each median time constant in $[0.5,500]$~h. Heating and cooling demand cannot fall in more severe weather, which the sign constraints impose on the first regime of the window, and the bound $\alpha\le1$ enforces the admissibility condition of Theorem~\ref{thm:fes}. Since $\mu^{\mathrm b}_t$ is linear in all parameters but $\bm\theta$, the next result shows that they can be fitted exactly for each $\bm\theta$.

\begin{proposition}[Exact inner problem]\label{prop:qp}
For fixed $\bm\theta$ and a design matrix of full column rank, \eqref{eq:qp} is a strictly convex quadratic program with a unique solution. Eliminating the unconstrained coefficients through the Schur complement of the Gram matrix leaves a nonnegative least-squares problem in the 48 sensitivities, which the Lawson--Hanson algorithm~\cite{lawson1974} solves in finitely many steps. The profile objective is continuous on the compact set $\Theta$, so a global minimizer $\hat{\bm\theta}$ exists.
\end{proposition}

We locate $\hat{\bm\theta}$ by a coordinate grid followed by a Nelder--Mead refinement~\cite{nelder1965}, and each evaluation reuses the Gram matrix of the calendar block. Gradient-boosted decision trees (GBDT) then fit the residuals of the backbone using the hour, weekday, day of year, holiday indicator, temperature, dew point, $F^{\HD}_t$, $F^{\CD}_t$ and the fractionally smoothed temperature, and the mean model is $\mu_t=\mu^{\mathrm b}_t+\mu^{\mathrm{gb}}_t$, where $\mu^{\mathrm{gb}}_t$ is this boosted correction. The backbone carries the level, the trend, the regimes and the thermal response, which must extrapolate beyond the training data, while the trees capture bounded interactions such as the effect of humidity. No load lag enters $\mu_t$, and short-term dynamics are left to the residual model of the next section, whose predictive law is exact.

\section{Long Memory, Breaks and Online Adaptation}\label{sec:memory}
\subsection{Multiscale Fractional Residual Model}\label{sec:resid}
Day-ahead residuals persist on three time scales, since errors within a day are correlated, errors at the same hour repeat across days and the daily level drifts. Assumption~\ref{as:resid} is therefore made specific by
\begin{equation}
y_t=X_t+P_{D(t),h(t)}+M_{D(t)},\label{eq:xpm}
\end{equation}
where $X$ is an hourly stationary fOU process, $dX_t=-\kappa X_t\,dt+\sigma\,dB^H_t$ with $B^H$ a fractional Brownian motion with Hurst exponent $H$, the 24 sequences $P_{\cdot,h}$ are independent daily fOU processes with common parameters, and $M$ is a daily fOU process. Hence
\begin{multline*}
\Cov(y_t,y_s)=\gamma_X(|t-s|)+\gamma_M(|D(t)-D(s)|)\\+\one\{h(t)=h(s)\}\,\gamma_P(|D(t)-D(s)|),
\end{multline*}
where each autocovariance $\gamma$ has the following closed form.

\begin{lemma}[Covariance of the fOU process]\label{lem:fou}
For $H\in(0,1)$ and $s\ge0$,
\begin{multline}
\gamma(s)=\frac{\sigma^2H}{2\kappa^{2H}}\Big[\Gamma(2H)e^{-\kappa s}+e^{\kappa s}\,\Gamma(2H,\kappa s)\\-\frac{(\kappa s)^{2H}}{2H}\,{}_1F_1(1;2H+1;-\kappa s)\Big],\label{eq:fouacov}
\end{multline}
where $\Gamma(\cdot,\cdot)$ is the upper incomplete gamma function and ${}_1F_1$ the confluent hypergeometric function. In particular $\gamma(0)=\sigma^2H\Gamma(2H)\kappa^{-2H}$, $\gamma(s)=\sigma^2e^{-\kappa s}/(2\kappa)$ for $H=\tfrac12$, and $\gamma(s)\sim\sigma^2H(2H-1)\kappa^{-2}s^{2H-2}$ as $s\to\infty$ for $H\neq\tfrac12$~\cite{cheridito2003}.
\end{lemma}

The scales $\vartheta_h$ are the hourly standard deviations of the residuals in a calibration year. The nine remaining parameters, a rate, a scale and a Hurst exponent for each component of \eqref{eq:xpm}, minimize the squared distance between the model covariance and the sample second moments of $y$ at lags of up to 14 days, within physical bounds that act as a regularization when the three scales are weakly separated. So that the model describes genuine out-of-sample errors, the calibration residuals for test year $Y$ are those of year $Y-1$ predicted by a mean model trained on data up to $Y-2$. Let $\bm r_p$ collect the 346 residuals from 00:00 of day $D-15$ to the issue time, and let $\bm\Sigma_{ff}$, $\bm\Sigma_{fp}$ and $\bm\Sigma_{pp}$ be the covariance blocks of the 24 residuals of day $D$ and of $\bm r_p$ implied by \eqref{eq:xpm}, \eqref{eq:fouacov} and the scales. The day-ahead law of $\bm L_D$ is Gaussian with
\begin{equation}
\hat{\bm\mu}_D=\bm\mu_D+\bm\Sigma_{fp}\bm\Sigma_{pp}^{-1}\bm r_p,\ \ \hat{\bm\Sigma}_D=\bm\Sigma_{ff}-\bm\Sigma_{fp}\bm\Sigma_{pp}^{-1}\bm\Sigma_{pf},\label{eq:predictive}
\end{equation}
where $\bm\mu_D$ holds the 24 mean forecasts of day $D$. The blocks depend only on the hours involved, so one Cholesky factorization of $\bm\Sigma_{pp}$ serves a whole year.

\subsection{Spurious Breaks and Joint Estimation}\label{sec:breaks}
The regimes $\mathcal R_j$ of \eqref{eq:backbone} are dated on daily means through the partial structural change model~\cite{bai2003}
\begin{equation}
\bar L_D=\bm f_D^\T\bm\zeta_{j(D)}+\tilde{\bm x}_D^\T\tilde{\bm\omega}+\xi_D,\label{eq:daily}
\end{equation}
where $\bar L_D$ is the mean load of day $D$ and $j(D)$ its regime. The vector $\bm f_D=(1,\bar F^{\HD}_D,\bar F^{\CD}_D)^\T$ holds a constant and the daily means of the filtered degree-hours, and its coefficients $\bm\zeta_j$, the level and sensitivities of regime $j$, are again sign-constrained. The weekday, holiday, seasonal and trend terms in $\tilde{\bm x}_D$ have coefficients $\tilde{\bm\omega}$ common to all regimes. The standard estimator, which combines least squares with the Bayesian information criterion (BIC), presumes that the errors $\xi_D$ have short memory, and the next result shows what happens when they do not.

\begin{proposition}[Memory-blind criteria find spurious breaks]\label{prop:spurious}
Let $\xi_D$ be stationary Gaussian with $\Cov(\xi_D,\xi_{D+k})\sim c_\xi k^{2d-1}$ for some $c_\xi>0$ and $d\in(0,\tfrac12)$, and let there be no break. For the mean-shift model with minimum regime length $\lfloor\rho n\rfloor$, $\rho\in(0,\tfrac12)$, the number of breaks $\hat m$ chosen by a BIC that treats $\xi_D$ as white noise satisfies $\Prob(\hat m\ge1)\to1$ as $n\to\infty$.
\end{proposition}

Under long memory, the partial sums of $\xi_D$ grow as $n^{d+1/2}$~\cite{taqqu1975}, so one spurious break reduces the residual sum of squares by an amount of order $n^{2d}$, whereas the BIC penalty grows only as $\log n$. Since level shifts in turn inflate estimates of $d$~\cite{diebold2001}, breaks and memory must be estimated together. Algorithm~\ref{alg:joint} minimizes the BIC of \eqref{eq:daily} after prewhitening with an autoregressive fractionally integrated moving average model, ARFIMA$(1,d,0)$. It alternates between the breaks, which dynamic programming finds exactly for a fixed memory~\cite{bai2003}, and the memory parameters, which minimize the conditional sum of squares for fixed breaks.

\begin{algorithm}[!t]
\small
\caption{Joint Break and Memory Estimation}\label{alg:joint}
\begin{algorithmic}[1]
\REQUIRE daily data $(\bar L_D,\bm f_D,\tilde{\bm x}_D)_{D=1}^{n}$, minimum regime length $\ell$, maximum number of breaks $m_{\max}$, grids $\mathcal G_d$ and $\mathcal G_\phi$
\STATE $(d,\phi)\leftarrow(0,0)$, $\tilde{\bm\omega}\leftarrow$ least squares without breaks, $\mathcal S\leftarrow\emptyset$
\REPEAT
\STATE Filter $\bar L_D$, $\bm f_D$, $\tilde{\bm x}_D$ by $(1-\phi B)(1-B)^d$, giving $\bar L^{w}_D$, $\bm f^{w}_D$, $\tilde{\bm x}^{w}_D$
\REPEAT
\STATE For segments with $j-i+1\ge\ell$, let $C(i,j)=\min_{\zeta_2,\zeta_3\ge0}\sum_{D=i}^{j}(\bar L^{w}_D-\tilde{\bm x}^{w\T}_D\tilde{\bm\omega}-\bm f^{w\T}_D\bm\zeta)^2$
\STATE For $m=0,\dots,m_{\max}$, find by dynamic programming the breaks that minimize $\mathrm{SSR}_m=\sum_{j=0}^{m}C(k_j+1,k_{j+1})$, with $k_0=0$ and $k_{m+1}=n$
\STATE $\hat m\leftarrow\arg\min_m\,n\log(\mathrm{SSR}_m/n)+[3(m+1)+m+\dim\tilde{\bm x}_D]\log n$, and refit $\tilde{\bm\omega}$ given the $\hat m$ breaks
\UNTIL{a set of breaks repeats}
\STATE $(d,\phi,\tilde{\bm\omega})\leftarrow$ minimizer of the conditional sum of squares of the filtered model \eqref{eq:daily} with these breaks, with $(d,\phi)\in\mathcal G_d\times\mathcal G_\phi$%
\STATE $\varpi\leftarrow(\hat m,k_1,\dots,k_{\hat m},d,\phi)$; stop if $\varpi\in\mathcal S$, else $\mathcal S\leftarrow\mathcal S\cup\{\varpi\}$
\UNTIL{the iteration limit is reached}
\ENSURE breaks $\hat k_1,\dots,\hat k_{\hat m}$, memory $(\hat d,\hat\phi)$, coefficients $\hat{\bm\zeta}_j$
\end{algorithmic}
\end{algorithm}

\begin{proposition}[Termination]\label{prop:termination}
Each block update of Algorithm~\ref{alg:joint} minimizes the criterion over its block, so the criterion never increases. Without an iteration limit, the algorithm stops after finitely many iterations at a point that no single block update improves.
\end{proposition}

Termination follows because breaks and memory parameters take finitely many values, so a tuple must repeat. The estimated breaks also set the training window of the mean model, which is the current regime extended backwards over earlier regimes until it covers at least 365 days. Regimes inside the window receive their own level and sensitivity changes in \eqref{eq:backbone}, and older data are discarded.

\subsection{Memory-Aware Online Detection}\label{sec:detector}
Between annual re-estimations, the level of demand can shift abruptly, as it did in March 2020. Let $\bar r_D$ be the mean residual of day $D$. Under Assumption~\ref{as:resid} it is stationary Gaussian with an autocovariance that follows from Lemma~\ref{lem:fou}. The Durbin--Levinson recursion gives the best linear predictor $\bar r_{D|D-1}=\sum_{j\ge1}\phi_{D,j}\bar r_{D-j}$ and the innovation $a_D=\bar r_D-\bar r_{D|D-1}$, with variance $\varsigma^2_D$. A level shift $\delta$ that starts on day $\nu$ adds $\delta\psi_{D,\nu}$ to $a_D$, where $\psi_{D,\nu}=1-\sum_{j=1}^{D-\nu}\phi_{D,j}$ is the signature of the shift in the innovations. With a prior $\delta\sim\mathcal N(0,\sigma_\delta^2)$, the mixture likelihood ratio of a change on day $k\le D$ is
\begin{equation}
\log\Lambda_D(k)=-\frac12\log\big(1+\sigma_\delta^2Q_D(k)\big)+\frac{\sigma_\delta^2Z_D(k)^2}{2[1+\sigma_\delta^2Q_D(k)]},\label{eq:glr}
\end{equation}
with $Z_D(k)=\sum_{i=k}^{D}\psi_{i,k}a_i/\varsigma^2_i$ and $Q_D(k)=\sum_{i=k}^{D}\psi^2_{i,k}/\varsigma^2_i$. The alarm day is $N_A=\min\{D:\max_{k\in\mathcal K_D}\log\Lambda_D(k)\ge\log A\}$, where the candidate set $\mathcal K_D=\{D-K+1,\dots,D-2\}$ covers a window of $K$ days and leaves at least three post-change days. At an alarm, the estimate $\hat\delta=Z_D(\hat k)/Q_D(\hat k)$ at the maximizing day $\hat k$ shifts the level of $\mu_t$ from day $\hat k$ on, and the detector restarts. Theorem~\ref{thm:detector} bounds its false alarms, its delay and its error before an alarm.

\begin{theorem}[Memory-aware detection]\label{thm:detector}
Let $\bar r_D$ be stationary Gaussian with known autocovariance, let $\Prob_\infty$ and $\Prob_\nu$ be its laws without a change and with a shift $\delta\neq0$ from day $\nu$, and let $\mathcal F_D$ be the information up to day $D$.
(i) For every autocovariance, window and set of skipped days, $\E_\infty[N_A]\ge A$.
(ii) If $\bar r_D$ is ARFIMA$(0,d,0)$ with $d\in[0,\tfrac12)$, innovation variance $\varsigma^2$ and an infinite past, then $\psi_{D,\nu}=\psi_{D-\nu}$ with $\psi_k=\frac{\Gamma(k+1-d)}{\Gamma(1-d)\Gamma(k+1)}\sim\frac{k^{-d}}{\Gamma(1-d)}$, and the information gathered $n$ days after the change is $I_n=\frac{\delta^2}{2\varsigma^2}\sum_{k<n}\psi_k^2$.
(iii) Under (ii), let $n_A=\min\{n:I_n\ge\log A\}$, so that $n_A\sim[2\varsigma^2(1-2d)\Gamma(1-d)^2\log A/\delta^2]^{1/(1-2d)}$ as $A\to\infty$. If $K\ge cn_A$ for a constant $c>1$, then $\sup_\nu\operatorname{ess\,sup}\Prob_\nu(N_A-\nu+1>(1+\epsilon)n_A\mid\mathcal F_{\nu-1},N_A\ge\nu)\to0$ for every $\epsilon>0$. If every past day is a candidate, $\sup_\nu\operatorname{ess\,sup}\E_\nu[(N_A-\nu+1)^+\mid\mathcal F_{\nu-1}]\le n_A(1+o(1))$.
(iv) If, in addition, $cn_A\le K\le Cn_A$ with $1<c\le C$, the excess squared one-step error accumulated up to the alarm, $\delta^2\sum_{k=0}^{N_A-\nu}\psi_k^2$, equals $2\varsigma^2\log A\,(1+o_P(1))$ on $\{N_A\ge\nu\}$, for every $d$ and $\delta$.
\end{theorem}

Part (i) holds because the statistic is dominated by the Shiryaev--Roberts statistic~\cite{tartakovsky2014}, whatever the memory. Part (ii) describes the self-healing of long-memory predictors, whose coefficients sum to one, so that a shift gradually enters the forecast and its footprint on the innovations fades as $k^{-d}$, whereas a short-memory predictor that reverts to a fixed mean keeps a permanent bias. Detection is therefore slower under memory, yet part (iv) shows that the error paid until the alarm does not depend on $d$. A detector that treats $\bar r_D$ as white noise ignores the inflation of the variance of partial sums, which grows as $K^{2d}$, and raises false alarms (Section~\ref{sec:mc}).

\begin{example}[Detection delay]\label{ex:delay}
With $A=1826$ days (five years) and $\delta=\varsigma$, $n_A$ is 16 days for $d=0$, 67 days for $d=0.2$ and 350 days for $d=0.3$, and for $\delta=2\varsigma$ it is 4, 8 and 14 days. In all cases the excess squared error at the alarm is about $2\log A\approx15$ innovation variances, and the window of $K=60$ days used in Section~\ref{sec:cases} covers shifts of two innovation standard deviations or more.
\end{example}

\section{Reserves From Joint Predictive Laws}\label{sec:reserves}
To schedule balancing reserves from the day-ahead law \eqref{eq:predictive}, let $e_h=L_h-\hat\mu_h$ be the imbalance between load and forecast at hour $h$ of day $D$. Under an affine real-time policy, the thermal fleet covers the share $1-b_h$ of the imbalance and a pumped-storage plant covers $b_he_h$. Thermal units are not limited in energy, but the plant must keep an energy headroom $E$ around its scheduled state of charge that absorbs the cumulative imbalance $\sum_{h\le n}b_he_h$ at every hour $n$ of the day. The next result shows how large $E$ must be when errors persist.

\begin{theorem}[Energy reserves under memory]\label{thm:energy}
Let $\bm e\sim\mathcal N(\bm0,\bm\Sigma)$ on $n$ hours, $V_n=\one^\T\bm\Sigma\one$, $\Vind_n=\operatorname{tr}\bm\Sigma$ and $z=\Phi^{-1}(1-\varepsilon/2)$, where $\Phi$ is the standard normal distribution function.
(i) The smallest $E$ with $\Prob(|\sum_te_t|\le E)\ge1-\varepsilon$ is $z\sqrt{V_n}$. Sizing $E$ from the diagonal of $\bm\Sigma$, as if the errors were independent, gives the violation probability $2\Phi(-z\sqrt{\Vind_n/V_n})$, which exceeds $\varepsilon$ if and only if the off-diagonal covariances have a positive sum.
(ii) If $e_t$ is a sampled stationary fOU process with $H\in(\frac12,1)$, then $V_n\sim\sigma^2\kappa^{-2}n^{2H}$, so the required headroom grows as $n^H$ and the violation probability of diagonal sizing tends to one. If $H=\frac12$, then $V_n/n\to\sigma^2\coth(\kappa/2)/(2\kappa)$.
(iii) Let $\bm\Sigma=\bm U^\T\bm U$. Under a Gaussian law, each chance constraint represented in \eqref{eq:socp} holds if and only if its second-order cone counterpart holds, so \eqref{eq:socp} is an exact convex reformulation.
\end{theorem}

Part (i) holds because $\sum_te_t$ is Gaussian with variance $V_n$, part (ii) follows from Lemma~\ref{lem:fou} by a Riemann-sum argument, and part (iii) applies part (i) to each constraint.  For the day-ahead laws \eqref{eq:predictive} of the case studies, $\sqrt{V_{24}/\Vind_{24}}$, the factor by which error correlations inflate the standard deviation of the daily error sum, averages \IsoCovRatio\ in ISO-NE and \PjmCovRatio\ in PJM. By part (i), a plant absorbing every imbalance with a headroom sized from hourly variances would fail on about \IsoCovViol\% and \PjmCovViol\% of days instead of 5\%. The realized errors confirm this, since their daily sum left the diagonal 95\% band on \IsoCovEmpDiag\% and \PjmCovEmpDiag\% of test days but the joint band on only \IsoCovEmpJoint\% and \PjmCovEmpJoint\%. The day-ahead reserve schedule therefore solves
\begin{equation}
\begin{aligned}
\min_{\bm b,\bm R,\bm S,E}\ &\textstyle\sum_h\pi^R_hR_h+\pi^S\sum_hS_h+\pi^EE\\
\text{s.t.}\ &R_h\ge z_P(1-b_h)\hat\sigma_h,\quad S_h\ge z_Pb_h\hat\sigma_h,\quad S_h\le\bar P,\\
&E\ge z_E\big\|\bm U(\bm b\odot\one_{1:n})\big\|_2,\quad n=1,\dots,24,\\
&0\le b_h\le1,\quad 0\le E\le\bar E/2,
\end{aligned}\label{eq:socp}
\end{equation}
where $\hat\sigma_h^2$ are the diagonal entries of $\hat{\bm\Sigma}_D=\bm U^\T\bm U$ in MW$^2$, $z_P$ and $z_E$ are the two-sided standard normal quantiles of the risk levels $\varepsilon_P$ and $\varepsilon_E$, $\odot$ is the elementwise product and $\one_{1:n}$ selects hours 1 to $n$. The first two constraints hold the hourly thermal and storage imbalances within $R_h$ and $S_h$ with probability $1-\varepsilon_P$, and the cone constraints hold the cumulative storage imbalance within $E$ with probability $1-\varepsilon_E$ at each hour. The headroom is symmetric around a mid-level state of charge, hence $E\le\bar E/2$, and its price $\pi^E=\max(\eta\max_h\pi^{\mathrm{DA}}_h-\min_h\pi^{\mathrm{DA}}_h,0)$ per MWh is the arbitrage margin that it forgoes, where $\eta$ is the round-trip efficiency and $\pi^{\mathrm{DA}}_h$ are the day-ahead (DA) energy prices. Algorithm~\ref{alg:online} combines the layers of Sections~\ref{sec:physics} to~\ref{sec:reserves} into the annual estimation and the daily operation of the framework.

\begin{algorithm}[!t]
\small
\caption{Annual Estimation and Day-Ahead Operation}\label{alg:online}
\begin{algorithmic}[1]
\REQUIRE hourly data up to the issue time, settings $\ell$, $K$, $A$, $\sigma_\delta$, $\varepsilon_P$, $\varepsilon_E$, prices and plant limits
\STATE \textit{Annual estimation on 1 January of test year $Y$.} Find $\hat{\bm\theta}$ by the profile search over $\Theta$ on the data up to $Y-2$, solving \eqref{eq:qp} exactly for each $\bm\theta$
\FOR{the calibration data (up to $Y-2$) and the final data (up to $Y-1$)}
\STATE Run Algorithm~\ref{alg:joint} on \eqref{eq:daily}, fit \eqref{eq:backbone}--\eqref{eq:qp} on the training window and add the boosted correction to obtain $\mu_t$
\ENDFOR
\STATE Standardize the calibration residuals of year $Y-1$ by $\vartheta_h$ and fit \eqref{eq:xpm} by covariance matching with \eqref{eq:fouacov}
\STATE Factorize $\bm\Sigma_{pp}$, compute the Durbin--Levinson predictor of $\bar r_D$ and set the level correction $\hat\Delta\leftarrow0$
\STATE \textit{Daily operation at 10:00 of day $D-1$, for each day $D$ of year $Y$.} Update \eqref{eq:glr} with $\bar r_{D-2}$ unless day $D-2$ lies within a day of a holiday. If $\log\Lambda_{D-2}(\hat k)\ge\log A$ at the maximizing $\hat k\in\mathcal K_{D-2}$, set $\hat\Delta\leftarrow\hat\Delta+\hat\delta$, subtract $\hat\delta$ from the residuals from day $\hat k$ on and restart the detector
\STATE Compute $\hat{\bm\mu}_D$ and $\hat{\bm\Sigma}_D$ by \eqref{eq:predictive} with mean $\bm\mu_D+\hat\Delta\one$ and the last 346 residuals
\STATE Solve \eqref{eq:socp} with $\hat{\bm\Sigma}_D$ and the prices of day $D$
\ENSURE law $\mathcal N(\hat{\bm\mu}_D,\hat{\bm\Sigma}_D)$ and schedule $(\bm b,\bm R,\bm S,E)$ for each day $D$
\end{algorithmic}
\end{algorithm}

\section{Case Studies}\label{sec:cases}
\subsection{Data, Benchmarks and Metrics}
The ISO-NE hourly system load, station-weighted temperature and dew point, day-ahead and real-time (RT) locational marginal prices and regulation clearing prices come from the standard market design files released with GEFCom2017~\cite{hong2019}, which also contain eight load zones. The PJM load, day-ahead prices and the weather of three stations come from the open data set of Ruan \emph{et al.}~\cite{ruan2020}. Table~\ref{tab:data} summarizes both data sets, whose test years, 2008 to 2016 for ISO-NE and 2019 to 2021 for PJM, cover a recession and the COVID-19 pandemic.

\begin{table}[!t]
\caption{Data and Test Design}\label{tab:data}
\centering\footnotesize
\setlength{\tabcolsep}{2.5pt}
\begin{tabular}{@{}lll@{}}
\toprule
 & ISO-NE & PJM \\
\midrule
Hourly data & \IsoPeriod & \PjmPeriod \\
Hours & \IsoNhours & \PjmNhours \\
Mean / peak load (GW) & \IsoMeanLoad\ / \IsoPeakLoad & \PjmMeanLoad\ / \PjmPeakLoad \\
Weather & weighted stations & three stations \\
Prices & DA, RT, regulation & DA \\
Test years & 2008--2016 & 2019--2021 \\
Storage plant & Northfield Mountain~\cite{ge2019} & Bath County~\cite{dominion,drax} \\
$\bar P$ / $\bar E$ & 1168 MW / 8.7 GWh & 3003 MW / 24 GWh \\
\bottomrule
\end{tabular}
\end{table}

The benchmarks cover the families reviewed in Section~\ref{sec:intro}, namely the vanilla multiple linear regression (MLR)~\cite{hong2010} and its recency extension~\cite{wang2016}, the Kalman-filtered adaptive generalized additive model (GAM)~\cite{obst2021,devilmarest2024}, updated every hour up to the issue time, adaptive probabilistic load forecasting (APLF)~\cite{alvarez2021}, GBDT without and with the load lags available at the issue time, a multilayer perceptron (MLP) with quantile outputs~\cite{mlp2025}, a long short-term memory (LSTM) network and a patch transformer representative of~\cite{forwardformer2024,transbilstm2025}, and a Kolmogorov--Arnold network (KAN)~\cite{kan2025}. All methods use the same information set and are retrained every year. The adaptive GAM uses its Kalman predictive variance and the MLP its quantile outputs. The regression, tree and APLF benchmarks receive hour-specific split-conformal intervals from the errors of year $Y-1$ under a model trained up to $Y-2$, as the proposed residual model does, and the LSTM, transformer and KAN use the errors on their held-out validation segment.

We report the mean absolute percentage error (MAPE), the root mean square error (RMSE), the pinball loss averaged over the quantiles $0.01,\dots,0.99$, and, for the central 90\% interval, the prediction interval coverage probability (PICP) and the Winkler score, which adds to the interval width twenty times the distance by which an outcome falls outside it. Diebold--Mariano tests with the Harvey--Leybourne--Newbold correction compare daily squared errors and daily Winkler scores.

\subsection{Thermal Physics}
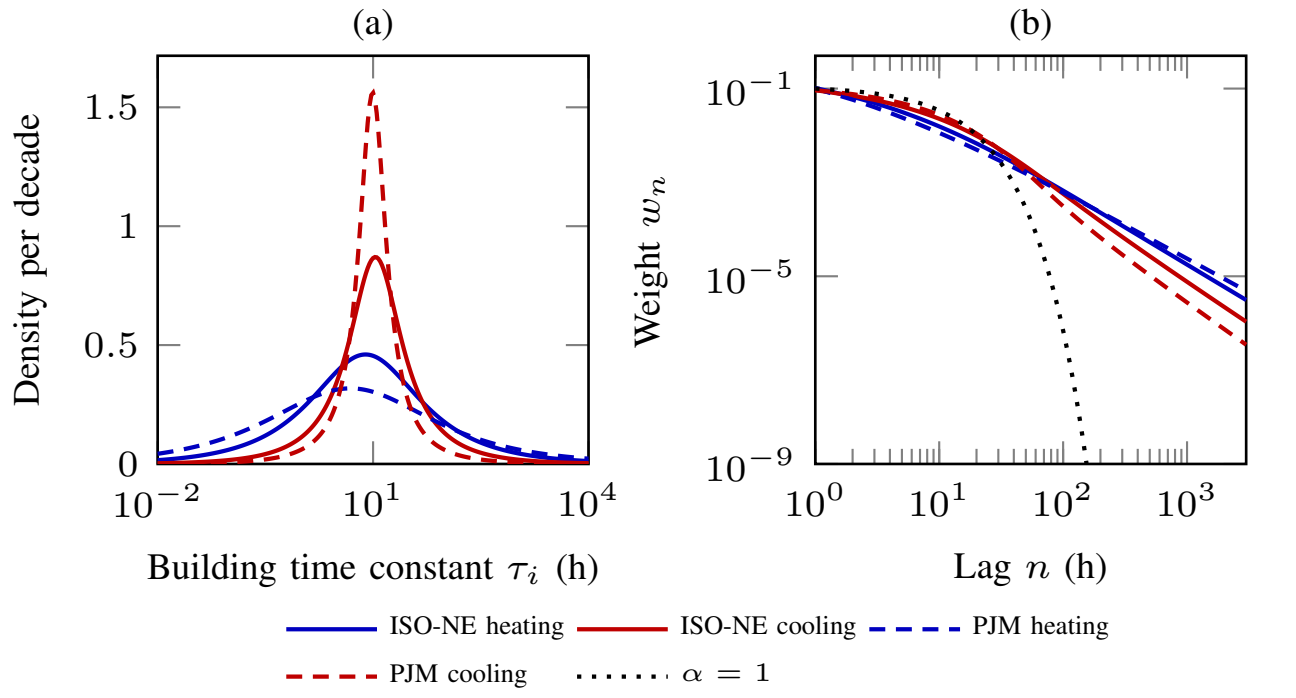
\begin{figure}[!t]
\centering
\begin{tikzpicture}
\begin{groupplot}[group style={group size=2 by 1, horizontal sep=15mm}, scale only axis, width=28.5mm, height=27mm]
\nextgroupplot[xmode=log, xmin=0.01, xmax=10000, ymin=0, xlabel={Building time constant $\tau_i$ (h)}, ylabel={Density per decade}, title={(a)}, title style={font=\scriptsize, yshift=-1.5ex}, legend to name=kernlegend, legend columns=3]
\addplot[isoH] table[x=tau, y=isoH] {data/colecole.dat}; \addlegendentry{ISO-NE heating}
\addplot[isoC] table[x=tau, y=isoC] {data/colecole.dat}; \addlegendentry{ISO-NE cooling}
\addplot[pjmH] table[x=tau, y=pjmH] {data/colecole.dat}; \addlegendentry{PJM heating}
\addplot[pjmC] table[x=tau, y=pjmC] {data/colecole.dat}; \addlegendentry{PJM cooling}
\addlegendimage{black, thick, dotted}\addlegendentry{$\alpha=1$}
\nextgroupplot[xmode=log, ymode=log, log basis y=10, xmin=1, xmax=3000, ymin=1e-9, ymax=0.5, xlabel={Lag $n$ (h)}, ylabel={Weight $w_n$}, title={(b)}, title style={font=\scriptsize, yshift=-1.5ex}]
\addplot[isoH] table[x=n, y=isoH] {data/kernel_weights.dat};
\addplot[isoC] table[x=n, y=isoC] {data/kernel_weights.dat};
\addplot[pjmH] table[x=n, y=pjmH] {data/kernel_weights.dat};
\addplot[pjmC] table[x=n, y=pjmC] {data/kernel_weights.dat};
\addplot[black, thick, dotted, restrict y to domain=-11:0] table[x=n, y=es] {data/kernel_weights.dat};
\end{groupplot}
\node at ($(group c1r1.south)!0.5!(group c2r1.south)+(0,-12.5mm)$) {\pgfplotslegendfromname{kernlegend}};
\end{tikzpicture}
\caption{Estimated thermal physics. (a) Dispersion of the building time constants implied by Proposition~\ref{prop:colecole}. (b) Kernel weights of \eqref{eq:fes}, which decay as $n^{-1-\alpha}$ (Theorem~\ref{thm:fes}), and the geometric decay of the exponential kernel ($\alpha=1$) with the heating time constant of ISO-NE.}
\label{fig:kernel}
\end{figure}

Fig.~\ref{fig:kernel} shows the estimated thermal physics over the full samples, where the heating order is $\hat\alpha_{\HD}=\IsoAlphaH$ in ISO-NE and \PjmAlphaH\ in PJM, with median time constants of \IsoTauH\ and \PjmTauH~h, while cooling is closer to a single time constant, with $\hat\alpha_{\CD}=\IsoAlphaC$ and \PjmAlphaC. Across the annual re-estimations, $\hat\alpha_{\HD}$ stays in $[\IsoAlphaHmin,\IsoAlphaHmax]$ in ISO-NE and in $[\PjmAlphaHmin,\PjmAlphaHmax]$ in PJM, far from the exponential case, and over the full samples the fractional kernels lower the residual sum of squares of the backbone by \IsoKernelSSERed\% and \PjmKernelSSERed\% relative to the best exponential kernels. By Proposition~\ref{prop:colecole}(c), the standard deviation of $\ln\tau_i$ for heating in ISO-NE is \IsoHSdLog, a factor of \IsoHSpread\ in the building time constant. The heating kernel puts \IsoHMassDay\% of its weight within the last day and still \IsoHMassWeek\% beyond one week, against \EsMassWeek\% for exponential smoothing with the same median time constant. Heating demand thus remembers cold spells for weeks, which no single smoothing constant can reproduce, while cooling demand has a shorter memory.

\subsection{Structural Breaks}
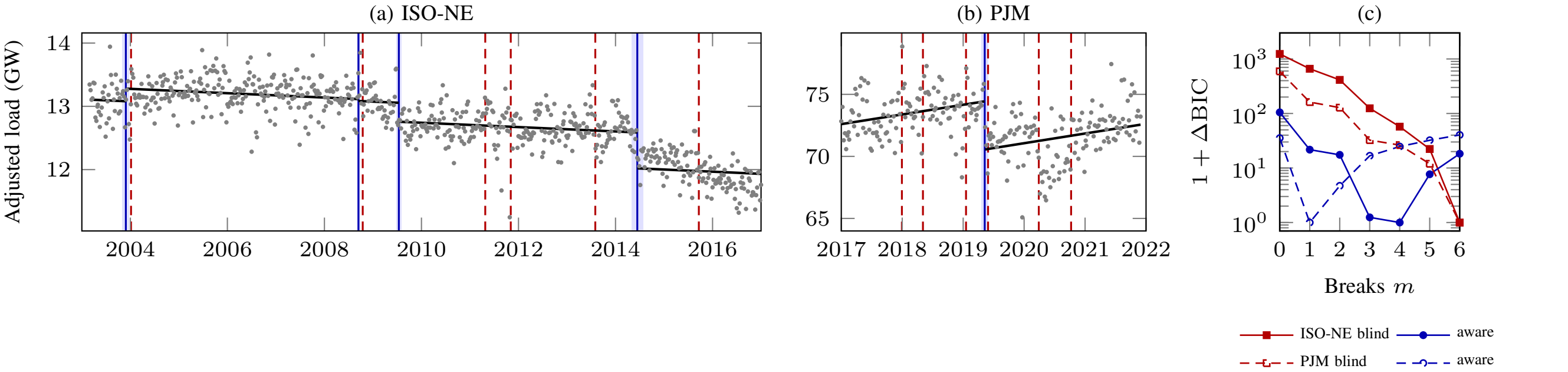
\begin{figure*}[!t]
\centering
\begin{tikzpicture}
\begin{axis}[name=bpiso, width=92mm, height=38mm, xmin=2003, xmax=2017, ymin=\IsoBPymin, ymax=\IsoBPymax, xtick={2004,2006,...,2016}, xticklabel style={/pgf/number format/1000 sep=}, ylabel={Adjusted load (GW)}, title={(a) ISO-NE}, title style={font=\scriptsize, yshift=-1.5ex}, set layers]
\begin{pgfonlayer}{axis background}
\fill[ciband] (axis cs:2003.8356,11.031) rectangle (axis cs:2003.9781,14.158);
\fill[ciband] (axis cs:2008.6721,11.031) rectangle (axis cs:2008.7322,14.158);
\fill[ciband] (axis cs:2009.4849,11.031) rectangle (axis cs:2009.5836,14.158);
\fill[ciband] (axis cs:2014.3288,11.031) rectangle (axis cs:2014.5699,14.158);
\draw[blindbreak] (axis cs:2004.0137,11.031) -- (axis cs:2004.0137,14.158);
\draw[blindbreak] (axis cs:2008.7896,11.031) -- (axis cs:2008.7896,14.158);
\draw[blindbreak] (axis cs:2011.3151,11.031) -- (axis cs:2011.3151,14.158);
\draw[blindbreak] (axis cs:2011.8411,11.031) -- (axis cs:2011.8411,14.158);
\draw[blindbreak] (axis cs:2013.5836,11.031) -- (axis cs:2013.5836,14.158);
\draw[blindbreak] (axis cs:2015.7178,11.031) -- (axis cs:2015.7178,14.158);
\draw[awarebreak] (axis cs:2003.9068,11.031) -- (axis cs:2003.9068,14.158);
\draw[awarebreak] (axis cs:2008.7022,11.031) -- (axis cs:2008.7022,14.158);
\draw[awarebreak] (axis cs:2009.5342,11.031) -- (axis cs:2009.5342,14.158);
\draw[awarebreak] (axis cs:2014.4493,11.031) -- (axis cs:2014.4493,14.158);\end{pgfonlayer}
\addplot[only marks, mark=*, mark size=0.45pt, gray] table[x=t, y=adj] {data/bp_isone_points.dat};
\addplot[black, thick, unbounded coords=jump] table[x=t, y=fit] {data/bp_isone_fit_aware.dat};
\end{axis}
\begin{axis}[name=bppjm, at={($(bpiso.east)+(9mm,0)$)}, anchor=west, width=50mm, height=38mm, xmin=2017, xmax=2022, ymin=\PjmBPymin, ymax=\PjmBPymax, xtick={2017,2018,...,2022}, xticklabel style={/pgf/number format/1000 sep=}, title={(b) PJM}, title style={font=\scriptsize, yshift=-1.5ex}, set layers]
\begin{pgfonlayer}{axis background}
\fill[ciband] (axis cs:2019.2959,63.986) rectangle (axis cs:2019.4110,79.953);
\draw[blindbreak] (axis cs:2017.9945,63.986) -- (axis cs:2017.9945,79.953);
\draw[blindbreak] (axis cs:2018.3397,63.986) -- (axis cs:2018.3397,79.953);
\draw[blindbreak] (axis cs:2019.0466,63.986) -- (axis cs:2019.0466,79.953);
\draw[blindbreak] (axis cs:2019.4082,63.986) -- (axis cs:2019.4082,79.953);
\draw[blindbreak] (axis cs:2020.2404,63.986) -- (axis cs:2020.2404,79.953);
\draw[blindbreak] (axis cs:2020.7705,63.986) -- (axis cs:2020.7705,79.953);
\draw[awarebreak] (axis cs:2019.3534,63.986) -- (axis cs:2019.3534,79.953);\end{pgfonlayer}
\addplot[only marks, mark=*, mark size=0.45pt, gray] table[x=t, y=adj] {data/bp_pjm_points.dat};
\addplot[black, thick, unbounded coords=jump] table[x=t, y=fit] {data/bp_pjm_fit_aware.dat};
\end{axis}
\begin{axis}[name=bic, at={($(bppjm.east)+(15mm,0)$)}, anchor=west, width=36mm, height=38mm, xmin=0, xmax=6, xtick={0,...,6}, ymode=log, ymin=0.7, ymax=3000, xlabel={Breaks $m$}, ylabel={$1+\Delta\mathrm{BIC}$}, title={(c)}, title style={font=\scriptsize, yshift=-1.5ex}, legend style={at={(0.5,-0.42)}, anchor=north, legend columns=2, font=\tiny}]
\addplot[red!70!black, mark=square*, mark size=1pt] table[x=m, y expr=\thisrow{isone_blind}+1] {data/bic.dat}; \addlegendentry{ISO-NE blind}
\addplot[blue!70!black, mark=*, mark size=1pt] table[x=m, y expr=\thisrow{isone_aware}+1] {data/bic.dat}; \addlegendentry{aware}
\addplot[red!70!black, densely dashed, mark=square, mark size=1pt] table[x=m, y expr=\thisrow{pjm_blind}+1] {data/bic.dat}; \addlegendentry{PJM blind}
\addplot[blue!70!black, densely dashed, mark=o, mark size=1pt] table[x=m, y expr=\thisrow{pjm_aware}+1] {data/bic.dat}; \addlegendentry{aware}
\end{axis}
\end{tikzpicture}
\caption{Bai--Perron graphs of the weather- and calendar-adjusted daily load (weekly means, gray) with the memory-aware regime fit (black), its break dates (blue lines) with 95\% confidence intervals (shaded) and the memory-blind break dates (red dashed) for (a) ISO-NE and (b) PJM. (c) Excess $\Delta\mathrm{BIC}$ of the information criterion over its minimum against the number of breaks, on a logarithmic scale. The memory-blind curves keep decreasing up to the largest admissible number of breaks.}
\label{fig:bp}
\end{figure*}

Fig.~\ref{fig:bp} and Table~\ref{tab:breaks} contrast memory-blind and memory-aware segmentation of \eqref{eq:daily} over the full samples, with minimum regimes of 180 days in ISO-NE and of 90 days in the shorter PJM sample (the annual re-estimations use 180 days for both). The memory-blind criterion keeps decreasing up to the largest admissible number of breaks, \IsoBlindM\ in ISO-NE and \PjmBlindM\ in PJM, as Proposition~\ref{prop:spurious} predicts. Algorithm~\ref{alg:joint} finds an interior minimum with \IsoAwareM\ breaks in ISO-NE ($\hat d=\IsoAwareD$, $\hat\phi=\IsoAwarePhi$), where three breaks give almost the same criterion value, and \PjmAwareM\ in PJM ($\hat d=\PjmAwareD$). The ISO-NE break of \IsoAwareBreakB\ has a 95\% confidence interval~\cite{bai1997} from \IsoAwareCIB, which contains the bankruptcy of Lehman Brothers, and the next break follows the trough of the recession in June 2009. Joint estimation also keeps memory and breaks apart, as the local Whittle estimate of the residual memory in ISO-NE shows. It is \IsoDnoBreaks\ without breaks, memory-blind breaks absorb almost all of it and leave $\IsoBlindDresid$, and the memory-aware breaks leave \IsoAwareDresid, the persistence that the residual model then describes.

\begin{table}[!t]
\caption{Memory-Aware Breaks With 95\% Confidence Intervals and Coefficient Changes}\label{tab:breaks}
\centering\footnotesize
\setlength{\tabcolsep}{3pt}
\begin{tabular}{@{}llcccc@{}}
\toprule
 & Break date & 95\% interval & $\Delta$level & $\Delta\beta^{\HD}$ & $\Delta\beta^{\CD}$\\
 & & (days) & (GW) & (GW/$^\circ$C) & (GW/$^\circ$C)\\
\midrule
ISO-NE & 2003-11-28 & $\pm$26 & 0.20 & 0.021 & 0.096 \\
 & 2008-09-14 & $\pm$11 & $-$0.03 & $-$0.023 & $-$0.328 \\
 & 2009-07-15 & $\pm$18 & $-$0.30 & 0.023 & 0.328 \\
 & 2014-06-14 & $\pm$44 & $-$0.57 & 0.023 & 0.045 \\
PJM & 2019-05-10 & $\pm$21 & $-$3.88 & 0.004 & 0.103 \\
 
\bottomrule
\end{tabular}
\end{table}

\subsection{Forecast Accuracy}
Table~\ref{tab:main} reports the accuracy over \IsoNdays\ ISO-NE and \PjmNdays\ PJM test days. The proposed method attains a MAPE of \IsoPropMAPE\% in ISO-NE and \PjmPropMAPE\% in PJM, \IsoGainMAPE\% and \PjmGainMAPE\% below the best benchmark (\IsoBestMAPEName\ and \PjmBestMAPEName, respectively). Its pinball loss is \IsoGainPin\% and \PjmGainPin\% lower than that of the best probabilistic benchmark, its 90\% intervals cover \IsoPropPICP\% and \PjmPropPICP\% of the outcomes, and its squared-error advantage is significant against \IsoDMsig\ of the \IsoDMtot\ benchmarks in ISO-NE and \PjmDMsig\ of \PjmDMtot\ in PJM. Year by year, it has the lowest MAPE in every ISO-NE test year and in 2020 and 2021 in PJM, whereas the LSTM is lower in 2019, when only two years of PJM data were available for training (supplementary Table~SII). The strongest alternatives, the networks and GBDT with load lags, all see recent loads, which indicates that residual persistence carries information. The proposed method extracts it through an exact predictive law, not lag features (Lemma~\ref{lem:fou} and \eqref{eq:predictive}).

\begin{table*}[!t]
\caption{Day-Ahead Accuracy on the Test Years. Bold Marks the Best Value of Each Column (PICP Closest to 90\%)}\label{tab:main}
\centering\footnotesize
\setlength{\tabcolsep}{5pt}
\begin{tabular}{@{}lcccccccccc@{}}
\toprule
 & \multicolumn{5}{c}{ISO-NE, 2008--2016} & \multicolumn{5}{c}{PJM, 2019--2021}\\
\cmidrule(lr){2-6}\cmidrule(l){7-11}
 & MAPE & RMSE & Pinball & PICP & Winkler & MAPE & RMSE & Pinball & PICP & Winkler\\
Method & (\%) & (GW) & (GW) & (\%) & (GW) & (\%) & (GW) & (GW) & (\%) & (GW)\\
\midrule
Vanilla MLR~\cite{hong2010} & 3.52 & 0.696$^{\dagger}$ & 0.175 & 86.7 & 2.860$^{\dagger}$ & 4.45 & 5.04$^{\dagger}$ & 1.632 & 73.8 & 26.06$^{\dagger}$ \\
Recency MLR~\cite{wang2016} & 2.89 & 0.565$^{\dagger}$ & 0.137 & 86.1 & 2.244$^{\dagger}$ & 3.64 & 4.03$^{\dagger}$ & 1.293 & 74.1 & 20.84$^{\dagger}$ \\
Adaptive GAM~\cite{obst2021,devilmarest2024} & 2.79 & 0.556$^{\dagger}$ & 0.153 & 94.7 & 2.586$^{\dagger}$ & 3.21 & 3.83$^{\dagger}$ & 1.054 & 90.9 & 17.34$^{\dagger}$ \\
APLF~\cite{alvarez2021} & 4.72 & 1.041$^{\dagger}$ & 0.266 & \textbf{89.2} & 4.721$^{\dagger}$ & 5.45 & 6.80$^{\dagger}$ & 1.819 & \textbf{89.9} & 29.05$^{\dagger}$ \\
GBDT & 3.33 & 0.608$^{\dagger}$ & 0.110 & 84.1 & 1.767$^{\dagger}$ & 2.91 & 3.44$^{\dagger}$ & 1.068 & 81.4 & 17.72$^{\dagger}$ \\
GBDT with load lags & 1.94 & 0.404$^{\dagger}$ & 0.097 & 87.8 & 1.592$^{\dagger}$ & 2.38 & 2.92$^{\dagger}$ & 0.817 & 88.9 & 13.68$^{\dagger}$ \\
MLP~\cite{mlp2025} & 1.88 & 0.398$^{\dagger}$ & 0.100 & 87.5 & 1.660$^{\dagger}$ & 2.44 & 2.95$^{\dagger}$ & 0.797 & 86.6 & 12.89$^{\dagger}$ \\
LSTM~\cite{transbilstm2025} & 1.85 & 0.384$^{\dagger}$ & 0.092 & 87.2 & 1.545$^{\dagger}$ & 2.44 & 2.87$^{\dagger}$ & 0.783 & 83.7 & 12.70$^{\dagger}$ \\
Patch transformer~\cite{forwardformer2024} & 1.92 & 0.397$^{\dagger}$ & 0.096 & 86.6 & 1.634$^{\dagger}$ & 2.63 & 3.06$^{\dagger}$ & 0.861 & 82.6 & 13.79$^{\dagger}$ \\
KAN~\cite{kan2025} & 2.09 & 0.423$^{\dagger}$ & 0.104 & 87.7 & 1.743$^{\dagger}$ & 2.94 & 3.44$^{\dagger}$ & 0.956 & 81.5 & 15.61$^{\dagger}$ \\
\midrule
\textbf{Proposed} & \textbf{1.56} & \textbf{0.338} & \textbf{0.086} & 91.9 & \textbf{1.493} & \textbf{2.13} & \textbf{2.65} & \textbf{0.712} & 91.6 & \textbf{11.70} \\
 
\bottomrule
\multicolumn{11}{@{}l}{\scriptsize $^\dagger$The proposed method is significantly better at the 5\% level (Diebold--Mariano test on daily squared errors for RMSE and on daily Winkler scores).}
\end{tabular}
\end{table*}

\subsection{Ablation}
Table~\ref{tab:ablation} removes or replaces one component at a time on all PJM test years and on four ISO-NE test years spread over the period, 2009, 2012, 2014 and 2016. The largest losses come from removing the residual model, the boosted correction or the thermal memory, with MAPE rising by \IsoAblResdMAPE\%, \IsoAblGbmdMAPE\% and \IsoAblMemdMAPE\% in ISO-NE and by \PjmAblResdMAPE\%, \PjmAblGbmdMAPE\% and \PjmAblMemdMAPE\% in PJM. Within the residual model, a single time scale raises MAPE by \IsoAblSingledMAPE\% and \PjmAblSingledMAPE\%, and an ARFIMA$(1,d,0)$ residual by \IsoAblArfimadMAPE\% and \PjmAblArfimadMAPE\%. Fractional Hurst exponents matter more in PJM, where the ordinary Ornstein--Uhlenbeck (OU) residual with $H=\frac12$ costs \PjmAblOudMAPE\%, than in ISO-NE, where it costs \IsoAblOudMAPE\%, and at the day-ahead horizon the multiscale model needs no affine correction of the kind used in the improved fOU model~\cite{nikseresht2025}. Memory-blind breaks cost \IsoAblBlinddMAPE\% and \PjmAblBlinddMAPE\%, because spurious regimes shorten the training window, whereas ignoring breaks costs \IsoAblBrkdMAPE\% and \PjmAblBrkdMAPE\%. The exponential kernel $\alpha=1$ raises MAPE by \IsoAblFracdMAPE\% and \PjmAblFracdMAPE\% even with boosting, and the kernel matters most where the physics stands alone, since without boosting $\alpha=1$ raises MAPE from \IsoAblGbmMAPE\% to \IsoAblGbmesMAPE\% in ISO-NE and from \PjmAblGbmMAPE\% to \PjmAblGbmesMAPE\% in PJM.

\begin{table}[!t]
\caption{Ablation of the Proposed Framework}\label{tab:ablation}
\centering\footnotesize
\setlength{\tabcolsep}{1.6pt}
\begin{tabular}{@{}lcccccc@{}}
\toprule
 & \multicolumn{3}{c}{ISO-NE} & \multicolumn{3}{c}{PJM}\\
\cmidrule(lr){2-4}\cmidrule(l){5-7}
Variant & MAPE & Pinball & Winkler & MAPE & Pinball & Winkler\\
\midrule
Proposed (full) & \textbf{1.53} & \textbf{0.082} & 1.389 & 2.13 & \textbf{0.712} & \textbf{11.70} \\
\midrule
Exponential kernel ($\alpha=1$) & 1.58 & 0.086 & 1.460 & 2.16 & 0.719 & 11.77 \\
No thermal memory & 2.09 & 0.114 & 1.936 & 2.74 & 0.904 & 14.73 \\
No structural breaks & 1.58 & 0.085 & 1.434 & 2.17 & 0.722 & 11.87 \\
Memory-blind breaks & 1.65 & 0.089 & 1.522 & 2.33 & 0.772 & 12.58 \\
No boosted correction & 2.57 & 0.136 & 2.171 & 2.65 & 0.859 & 13.52 \\
\quad and exponential kernel & 2.68 & 0.142 & 2.262 & 2.80 & 0.898 & 13.92 \\
No residual model & 2.36 & -- & -- & 3.50 & -- & -- \\
OU residual ($H=\frac12$) & 1.55 & 0.083 & \textbf{1.383} & 2.28 & 0.750 & 12.29 \\
ARFIMA$(1,d,0)$ residual & 1.81 & 0.094 & 1.532 & 2.26 & 0.746 & 12.11 \\
Affine-corrected fOU & 1.56 & 0.083 & 1.396 & 2.18 & 0.726 & 11.93 \\
Single-scale fOU & 1.81 & 0.094 & 1.508 & 2.26 & 0.755 & 12.45 \\
No online detector & \textbf{1.53} & \textbf{0.082} & 1.389 & \textbf{2.12} & \textbf{0.712} & 11.75 \\
Break-adjusted calibration & 1.55 & 0.083 & 1.402 & 2.16 & 0.719 & 11.71 \\
 
\bottomrule
\end{tabular}
\end{table}

\subsection{Online Detection of Shifts and Outages}
\begin{figure}[!t]
\centering
\begin{tikzpicture}
\begin{groupplot}[group style={group size=1 by 2, vertical sep=9mm}, width=88mm]
\nextgroupplot[height=32mm, xmin=0, xmax=28, ymin=45, ylabel={Load (GW)}, xtick={0,7,14,21,28}, xticklabels={9 Mar., 16 Mar., 23 Mar., 30 Mar., 6 Apr.}, title={(a)}, title style={font=\scriptsize, yshift=-1.5ex}, legend style={at={(0.5,0.02)}, anchor=south, legend columns=4}]
\addplot[name path=hi, draw=none, forget plot] table[x=t, y=hi] {data/forecast_pjm_covid.dat};
\addplot[name path=lo, draw=none, forget plot] table[x=t, y=lo] {data/forecast_pjm_covid.dat};
\addplot[blue!15] fill between[of=hi and lo]; \addlegendentry{90\% interval}
\addplot[black, line width=0.5pt] table[x=t, y=y] {data/forecast_pjm_covid.dat}; \addlegendentry{Load}
\addplot[blue!80!black, line width=0.5pt] table[x=t, y=fc] {data/forecast_pjm_covid.dat}; \addlegendentry{Proposed}
\addplot[red!70!black, densely dashed, line width=0.45pt] table[x=t, y=gbm] {data/forecast_pjm_covid.dat}; \addlegendentry{GBDT with lags}
\nextgroupplot[height=28mm, xmin=-31, xmax=396, xtick={0,91,182,274,366}, xticklabels={Jan. 2020, Apr., Jul., Oct., Jan. 2021}, ylabel={Statistic $\max_k\log\Lambda_D(k)$}, ymin=0, title={(b)}, title style={font=\scriptsize, yshift=-1.5ex}]
\addplot[black, line width=0.5pt] table[x=t, y=stat] {data/detector_pjm.dat};
\addplot[red!70!black, densely dashed, domain=-31:396] {ln(1826)};
\addplot[only marks, mark=triangle*, mark size=2pt, red!70!black] table[x=t, y expr=ln(1826)] {data/detector_pjm_alarms.dat};
\end{groupplot}
\end{tikzpicture}
\caption{PJM during COVID-19. (a) Hourly load, proposed forecasts with 90\% intervals and GBDT forecasts with load lags during the onset of the lockdowns. (b) Detector statistic, threshold $\log A$ (dashed) and alarms (triangles).}
\label{fig:covid}
\end{figure}
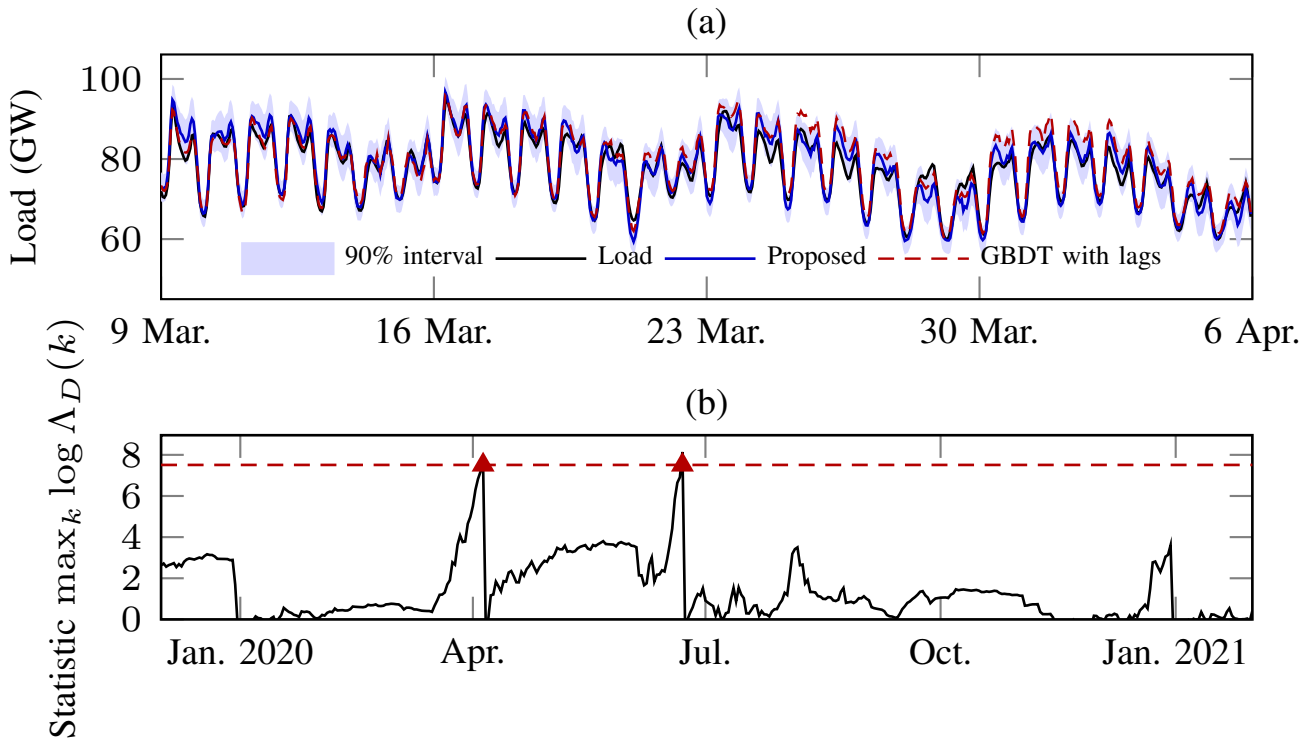

Fig.~\ref{fig:covid} follows PJM through 2020, when the detector raised an alarm on \PjmAlarmBDate\ that dated the shift to \PjmAlarmBStart, with $\hat\delta=\PjmAlarmBDelta$~GW, and a second alarm on \PjmAlarmCDate\ that dated the recovery to \PjmAlarmCStart, with $\hat\delta=\PjmAlarmCDelta$~GW. In ISO-NE, \IsoNalarmsEvent\ of the \IsoNalarms\ alarms raised over the \IsoNyears\ test years date a change to within days of a documented event (supplementary Table~SIII). These alarms come in pairs a few days apart around storms that cut power or halted activity across New England, namely the ice storm of December 2008, tropical storm Irene and the snowstorm of 2011, and the blizzard of January 2015, and the second alarm of each pair marks the recovery. The alarms thus date changes with a controlled false-alarm rate, but they hardly change accuracy (Table~\ref{tab:ablation}), because the self-healing of Theorem~\ref{thm:detector}(ii) already pulls the 14-day conditional law toward the new level. Their value lies in telling operators when and by how much demand has shifted.

\subsection{Multientity Benchmark on the GEFCom2017 Zones}
Adaptive multitask learning (AMTL)~\cite{zaballa2026} forecasts jointly the eight ISO-NE zones of GEFCom2017, five states and three areas of Massachusetts (NEMASSBOST, SEMASS and WCMASS), and reports its errors for 2016 over the 24 hours that follow 10:00 of each day (24~h). Table~\ref{tab:zones} compares the published figures with AMTL rerun under this protocol, where it improves on its published accuracy, and under our day-ahead protocol, whose 15 to 38~h horizons make it harder. Under the latter, the proposed method, applied zone by zone with zonal temperatures, averages a MAPE of \ZoneProp\%, against \ZoneAmtlDA\% for AMTL and \ZoneAmtlRep\% published for the easier protocol, and is more accurate than the published figures in \ZoneWins\ of the 8 zones.

\begin{table}[!t]
\caption{MAPE (\%) and RMSE (GW) on the GEFCom2017 Zones in 2016}\label{tab:zones}
\centering\scriptsize
\setlength{\tabcolsep}{3pt}
\begin{tabular}{@{}lccccccc@{}}
\toprule
 & \multicolumn{2}{c}{Published~\cite{zaballa2026}} & \multicolumn{2}{c}{AMTL rerun} & Proposed & \multicolumn{2}{c}{RMSE, DA}\\
\cmidrule(lr){2-3}\cmidrule(lr){4-5}\cmidrule(lr){6-6}\cmidrule(l){7-8}
Zone & APLF & AMTL & 24 h & DA & DA & AMTL & Proposed\\
\midrule
CT & 5.67 & 5.32 & 4.39 & 5.74 & \textbf{2.15} & 0.273 & \textbf{0.106} \\
ME & 3.63 & 3.40 & 2.88 & 3.61 & \textbf{2.21} & 0.060 & \textbf{0.038} \\
NEMASSBOST & 5.09 & 4.49 & 3.57 & 4.67 & \textbf{2.07} & 0.180 & \textbf{0.085} \\
NH & 4.62 & 4.19 & 3.42 & 4.34 & \textbf{1.87} & 0.078 & \textbf{0.035} \\
RI & 5.29 & 4.76 & 3.77 & 5.07 & \textbf{2.01} & 0.068 & \textbf{0.027} \\
SEMASS & 5.87 & 5.21 & 4.21 & 5.78 & \textbf{2.67} & 0.136 & \textbf{0.064} \\
VT & 5.11 & 4.84 & 4.05 & 5.36 & \textbf{3.31} & 0.044 & \textbf{0.028} \\
WCMASS & 5.32 & 4.70 & 3.81 & 5.04 & \textbf{2.14} & 0.127 & \textbf{0.054} \\
\midrule
Average & 5.08 & 4.61 & 3.76 & 4.95 & \textbf{2.30} & 0.121 & \textbf{0.055} \\
 
\bottomrule
\end{tabular}
\end{table}

\subsection{Value for Reserve Scheduling}
Table~\ref{tab:decision} schedules reserves with \eqref{eq:socp} for every test day of the years shown. Thermal reserve is priced at the ISO-NE regulation capacity clearing price and, in PJM, whose data lack regulation prices, at \PriceRatio\ times the day-ahead price, the ratio of the two averages in ISO-NE. The plant has $\eta=0.79$, the average round-trip efficiency of pumped storage in the United States~\cite{eia2021}, and $\pi^S=1$~\$/MW-h, shortfalls of power or energy cost \$1000/MWh, the ISO-NE reserve constraint penalty factor for 30-min operating reserve~\cite{george2014}, and $\varepsilon_P=\varepsilon_E=5\%$. Power shortfalls occur in \IsoDecViolP\% and \PjmDecViolP\% of hours, close to the design rate. Because the energy constraints are imposed hour by hour, the probability that the state of charge leaves its headroom at some hour of a day is not fixed in advance. The proposed law puts it at \IsoDecViolEModel\% and \PjmDecViolEModel\% on average, against realized rates of \IsoDecViolE\% and \PjmDecViolE\%. In PJM the realized rate is above the expected one even though the daily sum is calibrated, so the remaining gap lies in the intraday partial sums that the plant absorbs, a natural target for refinement. The same forecasts with a diagonal covariance expect \IsoDecViolEModelDiag\% and \PjmDecViolEModelDiag\% but fail on \IsoDecViolEDiag\% and \PjmDecViolEDiag\% of days, as Theorem~\ref{thm:energy} predicts, and so do all benchmarks, which supply marginal distributions only. The joint covariance buys more reserve, yet it lowers the total cost by \IsoDecSaveDiag\% and \PjmDecSaveDiag\%, and the proposed schedule costs \IsoDecSaveBase\% and \PjmDecSaveBase\% less than the cheapest benchmark schedule.

\begin{table}[!t]
\caption{Daily Reserve Cost and Violation Rates}\label{tab:decision}
\centering\footnotesize
\setlength{\tabcolsep}{2.4pt}
\begin{tabular}{@{}lcccccc@{}}
\toprule
 & \multicolumn{3}{c}{Cost (k\$/day)} & \multicolumn{3}{c}{Violations (\%)}\\
\cmidrule(lr){2-4}\cmidrule(l){5-7}
Forecast & Procure. & Penalty & Total & Power & Energy & Own law\\
\midrule
\multicolumn{7}{@{}l}{\textit{ISO-NE, Northfield Mountain (2015--2016)}} \\
Proposed (joint $\boldsymbol{\Sigma}$) & 135 & 219 & \textbf{355} & 6.5 & 4.1 & 3.2 \\
Proposed (diagonal) & 50 & 1272 & 1322 & 6.3 & 36.7 & 4.6 \\
GBDT with load lags & 52 & 2837 & 2888 & 11.8 & 46.8 & 4.7 \\
Adaptive GAM & 206 & 970 & 1176 & 1.5 & 25.2 & 4.7 \\
MLP & 57 & 1993 & 2050 & 10.4 & 41.5 & 4.7 \\
Recency MLR & 86 & 2955 & 3042 & 12.1 & 45.4 & 4.7 \\
\midrule
\multicolumn{7}{@{}l}{\textit{PJM, Bath County (2019--2021)}} \\
Proposed (joint $\boldsymbol{\Sigma}$) & 1583 & 2241 & \textbf{3824} & 5.1 & 8.5 & 5.3 \\
Proposed (diagonal) & 1113 & 10447 & 11560 & 5.1 & 57.5 & 7.6 \\
GBDT with load lags & 1209 & 11974 & 13183 & 5.9 & 57.8 & 7.6 \\
Adaptive GAM & 1807 & 10406 & 12213 & 5.7 & 53.1 & 7.6 \\
MLP & 1138 & 14151 & 15288 & 7.8 & 64.7 & 7.6 \\
Recency MLR & 1432 & 23642 & 25074 & 13.1 & 79.0 & 7.6 \\
 
\bottomrule
\multicolumn{7}{@{}p{\columnwidth}@{}}{\scriptsize Power violations are counted per hour and energy violations per day. The last column is the daily energy violation probability under the forecaster's own predictive law.}
\end{tabular}
\end{table}

\subsection{Monte Carlo Validation of the Theory}\label{sec:mc}
\begin{figure}[!t]
\centering
\begin{tikzpicture}
\begin{groupplot}[group style={group size=2 by 1, horizontal sep=10mm}, width=47mm, height=33mm]
\nextgroupplot[xmode=log, ymode=log, ymin=5, ymax=20000, xlabel={$\log A$}, ylabel={Mean delay (days)}, xtick={3,4,5,6,7,8}, xticklabels={3,4,5,6,7,8}, title={(a)}, title style={font=\scriptsize, yshift=-1.5ex}, legend style={at={(0.03,0.97)}, anchor=north west}, legend columns=3]
\addplot[black, mark=*, mark size=1pt, only marks] table[x=logA, y=d00] {data/sim_delay.dat}; \addlegendentry{$d=0$}
\addplot[blue!70!black, mark=square*, mark size=1pt, only marks] table[x=logA, y=d20] {data/sim_delay.dat}; \addlegendentry{$d=0.2$}
\addplot[red!70!black, mark=triangle*, mark size=1.3pt, only marks] table[x=logA, y=d30] {data/sim_delay.dat}; \addlegendentry{$d=0.3$}
\addplot[black, forget plot] table[x=logA, y=th00] {data/sim_delay.dat};
\addplot[blue!70!black, forget plot] table[x=logA, y=th20] {data/sim_delay.dat};
\addplot[red!70!black, forget plot] table[x=logA, y=th30] {data/sim_delay.dat};
\nextgroupplot[xlabel={Horizon $n$ (h)}, ylabel={$\sqrt{V_n/\Vind_n}$}, xmin=1, xmax=48, ymin=0, ymax=8, ytick={0,2,4,6,8}, title={(b)}, title style={font=\scriptsize, yshift=-1.5ex}, legend style={at={(0.5,0.98)}, anchor=north, /tikz/every even column/.append style={column sep=4pt}}, legend image post style={xscale=0.55}, legend columns=2]
\addplot[black] table[x=n, y=H50] {data/sim_energy.dat}; \addlegendentry{$H{=}0.5$}
\addplot[blue!70!black] table[x=n, y=H65] {data/sim_energy.dat}; \addlegendentry{$H{=}0.65$}
\addplot[red!70!black] table[x=n, y=H75] {data/sim_energy.dat}; \addlegendentry{$H{=}0.75$}
\addplot[orange!80!black] table[x=n, y=H85] {data/sim_energy.dat}; \addlegendentry{$H{=}0.85$}
\end{groupplot}
\end{tikzpicture}
\caption{Monte Carlo checks. (a) Mean delay of the detection of a shift $\delta=\varsigma$ (markers) and $n_A$ of Theorem~\ref{thm:detector}(iii) (lines). (b) Ratio of joint to independent standard deviations of cumulative errors (Theorem~\ref{thm:energy}), $\kappa=0.2$~h$^{-1}$.}
\label{fig:mc}
\end{figure}
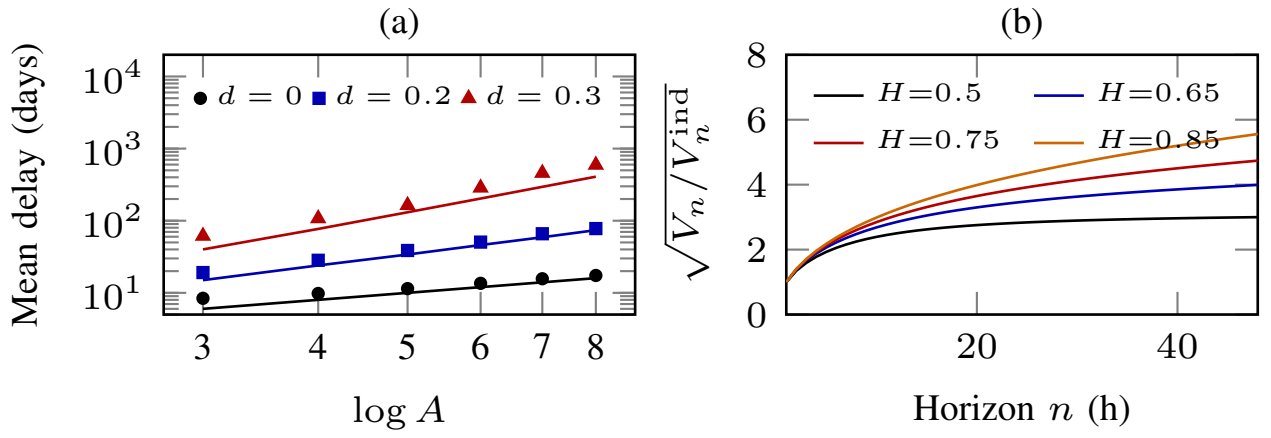

Fig.~\ref{fig:mc} checks Theorems~\ref{thm:detector} and~\ref{thm:energy} on simulated ARFIMA and fOU paths. Mean detection delays exceed $n_A$ of Theorem~\ref{thm:detector}(iii) by \SimDelayExcessMin\% to \SimDelayExcessMax\% at finite $A$ and grow with $\log A$ at the rate $1/(1-2d)$ predicts, with fitted log-log slopes of \SimSlopeZero, \SimSlopeTwo\ and \SimSlopeThree\ for $d=0$, $0.2$ and $0.3$ against the asymptotic values 1, 1.67 and 2.5. With $A=200$, the mean run length of the memory-aware detector never fell below \SimArlAwareMin\ days, whereas a memory-blind detector stopped after \SimArlBlindAtThree\ days on average for $d=0.3$. On long-memory series without breaks, memory-blind segmentation found \SimBlindBreaks\ breaks on average at $d=0.3$ against \SimAwareBreaks\ for Algorithm~\ref{alg:joint}, and on short-memory series with two level shifts the local Whittle estimate of $d$ rose to \SimDnaiveMax, while it stayed below \SimDjointMax\ after joint estimation.

\section{Conclusion}\label{sec:conclusion}
Day-ahead load forecasts and the reserves scheduled from them depend on the thermal inertia of buildings, structural breaks in demand and the long memory of forecast errors, which this paper has modeled jointly in one framework. The heterogeneity of the building stock makes the aggregate thermal response fractional, and its discrete form is a weighted average of past forcing exactly when the fractional order is at most one. Long memory makes memory-blind criteria report spurious breaks, so breaks and memory must be estimated together, and an online detector keeps its false-alarm guarantee under long memory, while the error accumulated before an alarm does not depend on the memory. Persistent errors also make the energy needed to absorb the cumulative error of a day grow faster than the square root of its length, so storage reserves sized from hourly variances fail with a probability that tends to one. On ISO-NE and PJM data, the framework lowered the MAPE of the best of ten benchmarks by \IsoGainMAPE\% and \PjmGainMAPE\%, dated the 2008 recession and the 2020 pandemic, and scheduled reserves that breached the storage energy limit on \IsoDecViolE\% and \PjmDecViolE\% of days, against \IsoDecViolEDiag\% and \PjmDecViolEDiag\% when the same forecasts ignored error covariance.

The framework carries over to other settings, as the break and detection layer applies to any forecast residual with long memory, Theorem~\ref{thm:energy} to any energy-limited resource, and the physics layer to any demand driven by first-order storage with dispersed time constants, such as heat pumps, water heaters, or district heating. The pipeline is also computationally lightweight enough for routine use. On a system equipped with an 8-core Intel Xeon processor and 12~GB of RAM, processing one test year takes \IsoMinPerYear~min for ISO-NE and \PjmMinPerYear~min for PJM, with most of the runtime devoted to annual estimation. The results hold under three assumptions that future work could relax, namely that temperatures are observed, that breaks shift only the level and the thermal sensitivities, and that the reserve program uses individual chance constraints with an affine policy. Relaxing them would let the residual model learn weather forecast errors, capture slow drifts beyond the trend and self-healing, and, through a scenario model built on the same predictive law, handle joint constraints and recourse.


\supplementlayout
\markboth{Supplementary Material}{Nikseresht: Supplementary Material}
\setcounter{section}{0}\setcounter{subsection}{0}\setcounter{equation}{0}\setcounter{table}{0}
\renewcommand{\theequation}{S\arabic{equation}}
\renewcommand{\thetable}{S\Roman{table}}
\renewcommand{\thesection}{S-\Roman{section}}

{\centering\vskip0.5\baselineskip\vskip0.2em{\Huge Supplementary Material\par}\par}
\vskip 2.5\baselineskip plus 0.75\baselineskip minus 0.375\baselineskip

\noindent This supplementary material completes the paper. Section~\ref{sec:proofs} proves every result, Section~\ref{sec:checks} checks the closed forms used in the proofs numerically, Section~\ref{sec:settings} lists the settings of the proposed framework and of the benchmarks, Section~\ref{sec:data} describes the data processing, and Section~\ref{sec:more} reports additional results. Equations, results, algorithms and tables numbered without the prefix S, such as (\ref{eq:fes}), Theorem~\ref{thm:fes} or Table~\ref{tab:main}, are those of the paper, whose notation is used throughout. Abbreviations are defined again at their first use.

\section{Proofs}\label{sec:proofs}

\subsection{Proposition~\ref{prop:colecole} (Aggregation)}
\textit{(a)} The Laplace transform of $g_\tau(s)=\tau^{-1}e^{-s/\tau}$ is $1/(1+p\tau)$, so Fubini's theorem~\cite{feller1971} gives, for $p>0$,
\begin{equation}
\int_0^\infty e^{-ps}g(s)\,ds=\int_{\mathbb R}\frac{f_\alpha(v)\,dv}{1+xe^{v}}=:G(x),\qquad x=p\bar\tau.\label{eq:G}
\end{equation}
Let $\Upsilon$ be a standard logistic variable independent of $V_i$, so that $\Prob(\Upsilon>t)=1/(1+e^t)$ and $\E[e^{i\omega\Upsilon}]=\pi\omega/\sinh(\pi\omega)$. For $x=e^y$, \eqref{eq:G} reads $G(e^y)=\Prob(\Upsilon>y+V_i)=\Prob(\Upsilon+V_i>y)$, because $V_i$ is symmetric, and the claimed transform is $1/(1+e^{\alpha y})=\Prob(\Upsilon/\alpha>y)$. It therefore suffices to show that $\Upsilon+V_i$ has the law of $\Upsilon/\alpha$. Integrating $e^{-i\kappa u}/(\cosh u+\cos\alpha\pi)$ along the boundary of the strip $0<\operatorname{Im}u<2\pi$, whose upper edge returns $e^{2\pi\kappa}$ times the integral over the real line and which contains the simple poles $i(\pi-\alpha\pi)$ and $i(\pi+\alpha\pi)$, gives
\begin{equation}
\int_{\mathbb R}\frac{e^{-i\kappa u}\,du}{\cosh u+\cos\alpha\pi}=\frac{2\pi\sinh(\alpha\pi\kappa)}{\sin(\alpha\pi)\sinh(\pi\kappa)},\nonumber
\end{equation}
and the substitution $u=\alpha v$ with $\kappa=\omega/\alpha$ yields $\E[e^{i\omega V_i}]=\sinh(\pi\omega)/[\alpha\sinh(\pi\omega/\alpha)]$. Hence $\E[e^{i\omega(\Upsilon+V_i)}]=\pi\omega/[\alpha\sinh(\pi\omega/\alpha)]=\E[e^{i\omega\Upsilon/\alpha}]$, and $G(x)=1/(1+x^\alpha)$ for $x>0$ by the uniqueness theorem for characteristic functions. The Caputo derivative of a function with a quiescent past has the Laplace transform $p^{\alpha}\hat\chi(p)$, and $\chi=g\ast u$ satisfies $[1+(\bar\tau p)^\alpha]\hat\chi=\hat u$, which is the transform of $\bar\tau^{\alpha}\,{}^{C}\!D^{\alpha}\chi+\chi=u$.

\textit{(b)} The series $E_\alpha(-t^\alpha)=\sum_{k\ge0}(-1)^kt^{\alpha k}/\Gamma(\alpha k+1)$ transforms termwise, for $p>1$, into $\sum_k(-1)^kp^{-\alpha k-1}=p^{\alpha-1}/(1+p^\alpha)$. Since $E_\alpha(0)=1$, the transform of $-\frac{d}{dt}E_\alpha(-t^\alpha)$ is $1-p^{\alpha}/(1+p^\alpha)=1/(1+p^\alpha)$ for $p>1$, and, by (a) and the uniqueness theorem for Laplace transforms~\cite{feller1971}, the scaling $t=s/\bar\tau$ identifies $g$. As a mixture of decreasing exponentials, $g$ is completely monotone, and $\int g=1$ follows from (a) as $p\downarrow0$. Finally, $1-\int_0^\infty e^{-ps}g(s)\,ds=(\bar\tau p)^\alpha+o(p^\alpha)$ as $p\downarrow0$, so Karamata's Tauberian theorem~\cite{feller1971} gives $\int_s^\infty g\sim(s/\bar\tau)^{-\alpha}/\Gamma(1-\alpha)$, and since $g$ is monotone, the monotone density theorem gives $g(s)\sim\alpha\bar\tau^{\alpha}s^{-1-\alpha}/\Gamma(1-\alpha)$.

\textit{(c)} The density $f_\alpha$ is even, so $\Prob(\tau_i\le\bar\tau)=\frac12$. With $a=\alpha\pi$ and $U=\alpha V_i$, the identities $\int_{\mathbb R}dx/(\cosh x+\cos a)=2a/\sin a$ and $\int_{\mathbb R}x^2dx/(\cosh x+\cos a)=2a(\pi^2-a^2)/(3\sin a)$, valid for $0<a<\pi$, give $\int f_\alpha=1$ and $\E[U^2]=\pi^2(1-\alpha^2)/3$, hence $\Var(V_i)=\pi^2(\alpha^{-2}-1)/3$. \hfill$\blacksquare$

\subsection{Theorem~\ref{thm:fes} (Fractional Exponential Smoothing)}
\textit{(i)} Let $\alpha\in(0,1)$ and $\bar\tau=c^{1/\alpha}$. By Proposition~\ref{prop:colecole}(a), $G(x)=1/(1+x^\alpha)$ for $x>0$. Both sides are analytic on the half-plane $\operatorname{Re}x>0$. On the left, $|1+xe^v|\ge1+e^v\operatorname{Re}x\ge1$, so the integrand of \eqref{eq:G} is bounded by $f_\alpha(v)$ and analytic in $x$. On the right, $|\arg x^\alpha|<\alpha\pi/2$, so $\operatorname{Re}x^\alpha>0$ and $1+x^\alpha\neq0$. By the identity theorem~\cite[Sec.~1.10]{olver2010}, the two sides agree on the half-plane. For $|z|<1$ the point $x=\bar\tau(1-z)$ has a positive real part and $x^\alpha=c(1-z)^\alpha$, so
\begin{equation}
W_{\alpha,c}(z)=\int_{\mathbb R}\frac{f_\alpha(v)\,dv}{1+\bar\tau e^{v}(1-z)}=\int_{\mathbb R}f_\alpha(v)\frac{\lambda_v}{1-(1-\lambda_v)z}\,dv,\qquad\lambda_v=\frac{1}{1+\bar\tau e^{v}},\nonumber
\end{equation}
because $1+\bar\tau e^v(1-z)=[1-(1-\lambda_v)z]/\lambda_v$. For $z\in[0,1)$, expanding $\lambda_v/[1-(1-\lambda_v)z]=\sum_n\lambda_v(1-\lambda_v)^nz^n$ and exchanging sum and integral by Tonelli's theorem~\cite{feller1971} gives $W_{\alpha,c}(z)=\sum_nz^n\E[\lambda_V(1-\lambda_V)^n]$. Power series that agree on $[0,1)$ have equal coefficients, so $w_n=\E[\lambda_V(1-\lambda_V)^n]$. Since $\lambda_V\in(0,1)$, every $w_n$ is positive and $\sum_nw_n=\E[\lambda_V\sum_n(1-\lambda_V)^n]=1$. Moreover, $(-\Delta)^kw_n=\E[\lambda_V^{k+1}(1-\lambda_V)^n]\ge0$ for all $k,n\ge0$, where $\Delta w_n=w_{n+1}-w_n$, which is the complete monotonicity of $(w_n)$. For $\alpha=1$, $W(z)=1/[1+c(1-z)]=\lambda/[1-(1-\lambda)z]$ with $\lambda=1/(1+c)$, hence $w_n=\lambda(1-\lambda)^n$.

For $\alpha\le1$ the weights also have a probabilistic reading. Write $1+c(1-z)^\alpha=(1+c)[1-q\,s(z)]$ with $q=c/(1+c)$ and $s(z)=1-(1-z)^\alpha=\sum_{j\ge1}s_jz^j$, where $s_j=(-1)^{j+1}\binom{\alpha}{j}=\alpha(1-\alpha)\cdots(j-1-\alpha)/j!\ge0$ and $\sum_js_j=s(1)=1$, so that $s$ is the probability generating function of the Sibuya law~\cite{sibuya1979}. Then $W(z)=(1-q)/[1-q\,s(z)]=\E[s(z)^G]$ for a geometric variable $G$ with $\Prob(G=k)=(1-q)q^k$, so $w_n=\Prob(S_1+\dots+S_G=n)$ for independent Sibuya variables $S_i$ that are independent of $G$. The lag of the aggregate response is thus a geometric number of Sibuya-distributed delays.

\textit{(ii)} Let $1<\alpha<2$. For $|z|<1$, $|\arg(1-z)|<\pi/2$ and hence $|\arg(1-z)^\alpha|<\alpha\pi/2<\pi$, so $1+c(1-z)^\alpha\neq0$. Thus $W$ is analytic in the unit disk and equals its power series there. Suppose that $w_n\ge0$ for all $n$. By Abel's theorem~\cite{feller1971} and monotone convergence, $\sum_nw_n=\lim_{z\uparrow1}W(z)=1$, so $(w_n)$ is a probability law, and its mean is $\sum_nnw_n=\lim_{z\uparrow1}W'(z)=\lim_{z\uparrow1}c\alpha(1-z)^{\alpha-1}[1+c(1-z)^\alpha]^{-2}=0$. A law on $\{0,1,\dots\}$ with zero mean is the point mass at $0$, so $W\equiv w_0=1/(1+c)<1$, which contradicts $W(1^-)=1$.

\textit{(iii)} For $|z|<1$, $\sum_{n\ge0}\bar w_nz^n=[1-W(z)]/(1-z)=c(1-z)^{\alpha-1}/[1+c(1-z)^\alpha]$, which behaves as $c(1-z)^{-(1-\alpha)}$ as $z\uparrow1$. The sequence $\bar w_n$ is nonincreasing, so Karamata's Tauberian theorem for power series with monotone coefficients~\cite[Sec.~XIII.5]{feller1971} gives $\bar w_n\sim c\,n^{-\alpha}/\Gamma(1-\alpha)$. By (i) the weights are nonincreasing. For $b>1$, $\bar w_n-\bar w_{\lfloor bn\rfloor}=\sum_{n<k\le\lfloor bn\rfloor}w_k\le(\lfloor bn\rfloor-n)w_n$, so
\begin{equation}
\liminf_{n\to\infty}n^{1+\alpha}w_n\ge\frac{c(1-b^{-\alpha})}{(b-1)\Gamma(1-\alpha)}\to\frac{\alpha c}{\Gamma(1-\alpha)}\quad(b\downarrow1),\nonumber
\end{equation}
and for $a\in(0,1)$, $\bar w_{\lfloor an\rfloor}-\bar w_n\ge(n-\lfloor an\rfloor)w_n$ gives $\limsup n^{1+\alpha}w_n\le c(a^{-\alpha}-1)/[(1-a)\Gamma(1-\alpha)]$, which tends to $\alpha c/\Gamma(1-\alpha)$ as $a\uparrow1$. Hence $w_n\sim\alpha c\,n^{-1-\alpha}/\Gamma(1-\alpha)$. The mean lag is $\sum_nnw_n=\sum_{n\ge0}\bar w_n=\infty$ because $\alpha<1$. For $\alpha=1$, $\bar w_n=\sum_{k>n}\lambda(1-\lambda)^k=(1-\lambda)^{n+1}$. \hfill$\blacksquare$

\textit{Example~\ref{ex:catalan}.} For $\alpha=\frac12$ and $c=1$, $(1+\sqrt{1-z})(1-\sqrt{1-z})=z$, so $W(z)=(1-\sqrt{1-z})/z$. The expansion $1-\sqrt{1-4y}=\sum_{n\ge0}2C_ny^{n+1}$ with $y=z/4$ gives $w_n=C_n/(2\cdot4^n)$, so that $w_0,\dots,w_7$ are 0.5, 0.125, 0.0625, 0.0391, 0.0273, 0.0205, 0.0161 and 0.0131, whose sum is 0.8036. Exponential smoothing with $\lambda=\frac12$ has the partial sum $1-2^{-8}=0.9961$. Stirling's formula~\cite{olver2010} gives $C_n\sim4^n/(\sqrt\pi n^{3/2})$, hence $w_n\sim n^{-3/2}/(2\sqrt\pi)$ and $\bar w_n\sim n^{-1/2}/\sqrt\pi$, in agreement with part (iii).

\subsection{Lemma~\ref{lem:fou} (Covariance of the Fractional Ornstein--Uhlenbeck Process)}
Let $H\in(\frac12,1)$. The stationary solution is $X_t=\sigma\int_{-\infty}^te^{-\kappa(t-u)}dB^H_u$, and for deterministic integrands $f$ and $g$ with $\iint|f(u)||g(v)||u-v|^{2H-2}du\,dv<\infty$, $\E[\int f\,dB^H\int g\,dB^H]=H(2H-1)\iint f(u)g(v)|u-v|^{2H-2}du\,dv$. With $f(u)=e^{\kappa u}\one\{u\le0\}$, $g(v)=e^{-\kappa(s-v)}\one\{v\le s\}$ and the substitutions $a=-u$ and $b=s-v$,
\begin{equation}
\gamma(s)=\sigma^2H(2H-1)\int_0^\infty\!\!\int_0^\infty e^{-\kappa(a+b)}|b-a-s|^{2H-2}\,da\,db.\nonumber
\end{equation}
If $\mathcal A$ and $\mathcal B$ are independent exponential variables with rate $\kappa$, then $\mathcal B-\mathcal A$ has the density $\frac\kappa2e^{-\kappa|q|}$, so
\begin{equation}
\gamma(s)=\frac{\sigma^2H(2H-1)}{2\kappa}\int_{\mathbb R}e^{-\kappa|q|}|q-s|^{2H-2}dq=\frac{\sigma^2H(2H-1)}{2\kappa}(J_1+J_2+J_3),\nonumber
\end{equation}
where, for $s\ge0$, the integrals over $(-\infty,0)$, $(0,s)$ and $(s,\infty)$ are
\begin{align*}
J_1&=e^{\kappa s}\kappa^{1-2H}\Gamma(2H-1,\kappa s),\qquad J_3=e^{-\kappa s}\kappa^{1-2H}\Gamma(2H-1),\\
J_2&=e^{-\kappa s}\int_0^se^{\kappa y}y^{2H-2}dy=\frac{s^{2H-1}}{2H-1}\,{}_1F_1(1;2H;-\kappa s),
\end{align*}
the last step by the integral representation of ${}_1F_1(2H-1;2H;\kappa s)$ and Kummer's transformation ${}_1F_1(a;b;x)=e^x{}_1F_1(b-a;b;-x)$~\cite{olver2010}. The identities $(2H-1)\Gamma(2H-1)=\Gamma(2H)$, $(2H-1)\Gamma(2H-1,x)=\Gamma(2H,x)-x^{2H-1}e^{-x}$ and $x^{2H-1}[{}_1F_1(1;2H;-x)-1]=-\frac{x^{2H}}{2H}{}_1F_1(1;2H+1;-x)$, the last of which follows from $(2H)_{k+1}=2H\,(2H+1)_k$, collect the terms into (\ref{eq:fouacov}).

For general $H\in(0,1)$, the spectral representation~\cite{cheridito2003} gives $\gamma(s)=\sigma^2\frac{\Gamma(2H+1)\sin(\pi H)}{2\pi}\int_{\mathbb R}e^{isx}\frac{|x|^{1-2H}}{\kappa^2+x^2}dx$ for $s>0$. For fixed $s>0$ this integral is real-analytic in $H\in(0,1)$, being the cosine transform of a function that is integrable at the origin and at infinity uniformly on compact subsets of $(0,1)$. The right side of (\ref{eq:fouacov}) is real-analytic in $H$ as a composition of analytic functions. The two agree on $(\frac12,1)$, hence on $(0,1)$ by the identity theorem for real-analytic functions~\cite{krantz2002}, and at $s=0$ by continuity. For $H=\frac12$, ${}_1F_1(1;2;-x)=(1-e^{-x})/x$ and $\Gamma(1,x)=e^{-x}$ give $\gamma(s)=\sigma^2e^{-\kappa s}/(2\kappa)$. The behavior for large $s$ follows from $\Gamma(2H,x)\sim x^{2H-1}e^{-x}$ and the asymptotic expansion of ${}_1F_1(1;2H+1;-x)$~\cite{cheridito2003}. \hfill$\blacksquare$

\subsection{Proposition~\ref{prop:qp} (Exact Inner Problem)}
For fixed $\bm\theta$ the objective is $\|\bm L-\bm X\bm\beta\|^2=\bm\beta^\T\bm G\bm\beta-2\bm g^\T\bm\beta+\|\bm L\|^2$, where $\bm X$ is the design, $\bm\beta$ stacks all linear coefficients and $\bm G=\bm X^\T\bm X\succ0$ by the rank condition, assumed for every $\bm\theta\in\Theta$. The problem is therefore strictly convex with a unique minimizer over the closed convex feasible set. Partition $\bm\beta=(\bm\beta_f,\bm\beta_c)$ into free and sign-constrained parts. For fixed $\bm\beta_c$, the optimal free part is $\bm\beta_f=\bm G_{ff}^{-1}(\bm g_f-\bm G_{fc}\bm\beta_c)$, and substituting it leaves $\bm\beta_c^\T\bm S\bm\beta_c-2\bm h^\T\bm\beta_c+\text{const}$ with the Schur complement $\bm S=\bm G_{cc}-\bm G_{cf}\bm G_{ff}^{-1}\bm G_{fc}\succ0$ and $\bm h=\bm g_c-\bm G_{cf}\bm G_{ff}^{-1}\bm g_f$. With $\bm S=\bm C\bm C^\T$, this equals $\|\bm C^\T\bm\beta_c-\bm C^{-1}\bm h\|^2+\text{const}$, a nonnegative least-squares problem in the 48 sensitivities whose active-set algorithm terminates with the exact solution after finitely many steps~\cite{lawson1974}. The profile $\mathrm{SSE}(\bm\theta)$ is the minimum of a jointly continuous function over a feasible set that does not depend on $\bm\theta$. The unique minimizer depends continuously on $(\bm G,\bm g)$, which are continuous in $\bm\theta$ because the weights $w_n$ of (\ref{eq:fes}) are continuous in $(\alpha,c)$ and the filter is a finite convolution with an exact unit-gain correction for the truncated tail (Section~\ref{sec:settings}). Hence $\mathrm{SSE}$ is continuous on the compact set $\Theta$ and attains its minimum. \hfill$\blacksquare$

\subsection{Proposition~\ref{prop:spurious} (Spurious Breaks)}
For the mean-shift model, the Bayesian information criterion (BIC) of $m$ breaks is $\mathrm{BIC}(m)=n\log(\mathrm{SSR}_m/n)+p_m\log n$, where $\mathrm{SSR}_m$ is the smallest residual sum of squares with $m$ breaks and $p_m$ the number of parameters, so that one break adds a fixed number $k_b=p_1-p_0$ of parameters. Let $\Sigma_\xi(r)=\sum_{D\le\lfloor rn\rfloor}\xi_D$. For one break after $\lfloor\lambda n\rfloor$ days, the reduction of the residual sum of squares relative to no break is
\begin{equation}
\Delta(\lambda)=n\lambda(1-\lambda)(\bar\xi_1-\bar\xi_2)^2=\frac{[\Sigma_\xi(\lambda)-\lambda\Sigma_\xi(1)]^2}{n\lambda(1-\lambda)},\nonumber
\end{equation}
up to the rounding of $\lambda n$, where $\bar\xi_1$ and $\bar\xi_2$ are the means of the two regimes. Since $\Var\Sigma_\xi(1)\sim c_\xi n^{2d+1}/[d(2d+1)]$, the functional limit theorem for Gaussian sequences with regularly varying covariances~\cite{taqqu1975} gives $n^{-(d+1/2)}\Sigma_\xi(\cdot)\Rightarrow c'B_H(\cdot)$ in $D[0,1]$, with $H=d+\frac12$ and $c'^2=c_\xi/[d(2d+1)]$. By the continuous mapping theorem,
\begin{equation}
n^{-2d}\sup_{\lambda\in[\rho,1-\rho]}\Delta(\lambda)\Rightarrow\Xi=c'^2\sup_{\lambda\in[\rho,1-\rho]}\frac{[B_H(\lambda)-\lambda B_H(1)]^2}{\lambda(1-\lambda)},\nonumber
\end{equation}
and $\Xi>0$ almost surely because the fractional Brownian bridge is not identically zero on $[\rho,1-\rho]$. The residual variance without a break satisfies $\mathrm{SSR}_0/n\to\gamma_\xi(0)$ in probability by the ergodic theorem~\cite{samorodnitsky2016}, which applies because a stationary Gaussian sequence whose covariances vanish at infinity is ergodic. With the optimal reduction $\Delta^*$ of one break, $\mathrm{BIC}(1)-\mathrm{BIC}(0)=n\log(1-\Delta^*/\mathrm{SSR}_0)+k_b\log n\le-n\Delta^*/\mathrm{SSR}_0+k_b\log n$. By the two limits above, $n^{1-2d}\Delta^*/\mathrm{SSR}_0\Rightarrow\Xi/\gamma_\xi(0)$, which is positive almost surely, and $n^{2d}/\log n\to\infty$ because $d>0$, so the right side, $-n^{2d}\,(n^{1-2d}\Delta^*/\mathrm{SSR}_0)+k_b\log n$, tends to $-\infty$ in probability. Therefore $\Prob(\mathrm{BIC}(1)<\mathrm{BIC}(0))\to1$, and since $\hat m$ minimizes the criterion over a set that contains $m=1$, $\Prob(\hat m=0)\to0$. \hfill$\blacksquare$

The same argument applies to the partial structural change model (\ref{eq:daily}) after projecting out the common regressors, because projecting a long-memory sequence on a fixed number of smooth regressors keeps the $n^{d+1/2}$ scaling of its partial sums. A fixed minimum regime length $\ell$, as used in the paper, only enlarges the set of admissible break dates and makes spurious breaks more likely.

\subsection{Proposition~\ref{prop:termination} (Termination of Algorithm~\ref{alg:joint})}
For a set of breaks $\mathcal K$ and memory parameters $(d,\phi)$, let $\mathrm{CSS}(\mathcal K,\tilde{\bm\omega},d,\phi)$ be the smallest conditional sum of squares of the filtered model (\ref{eq:daily}) over the regime coefficients $\bm\zeta_j$, with nonnegative sensitivities, and let
\begin{equation}
\tilde J(\mathcal K,\tilde{\bm\omega},d,\phi)=n\log\frac{\mathrm{CSS}(\mathcal K,\tilde{\bm\omega},d,\phi)}{n}+P(|\mathcal K|),\qquad J(\mathcal K,d,\phi)=\min_{\tilde{\bm\omega}}\tilde J(\mathcal K,\tilde{\bm\omega},d,\phi),\nonumber
\end{equation}
where $P(m)=[3(m+1)+m+\dim\tilde{\bm x}_D]\log n$ is the penalty of the information criterion in line 7 of Algorithm~\ref{alg:joint}. For fixed $(\tilde{\bm\omega},d,\phi)$, the segment costs $C(i,j)$ of line 5 are additive over segments, so dynamic programming returns the exact minimum of the residual sum of squares for every $m$~\cite{bai2003}, and line 7 returns a global minimizer of $\tilde J$ over $\mathcal K$. The refit in line 7 minimizes $\tilde J$ over $\tilde{\bm\omega}$ for the new breaks, so $\tilde J$ never increases in the inner loop. The memory update of line 9 minimizes $J(\mathcal K,\cdot,\cdot)$ over the grid $\mathcal G_d\times\mathcal G_\phi$, because the penalty does not depend on $(d,\phi)$, and sets $\tilde{\bm\omega}$ to the corresponding minimizer. Let $\varpi_t=(\mathcal K_t,d_t,\phi_t)$ be the tuple of outer iteration $t$. After line 9 the state of the algorithm is a function of $\varpi_t$, and its criterion is $\tilde J=J(\varpi_t)$. The inner loop of iteration $t+1$ starts from this state, so it ends with breaks $\mathcal K_{t+1}$ such that $J(\mathcal K_{t+1},d_t,\phi_t)\le J(\varpi_t)$, and line 9 then gives $J(\varpi_{t+1})\le J(\mathcal K_{t+1},d_t,\phi_t)$. The criterion therefore never increases.

For given $n$ and $\ell$ the sets of breaks are finitely many, and so are the grid points, so the inner loop stops when a set of breaks repeats, and the tuples $\varpi_t$ must repeat as well. Without an iteration limit, the algorithm stops at the first repetition, $\varpi_t=\varpi_s$ with $s<t$, and monotonicity gives $J(\varpi_s)=J(\varpi_{s+1})=\dots=J(\varpi_t)$. Because the state is a function of the tuple, a further iteration would repeat iteration $s+1$, in which no update of the breaks lowered the criterion. The refit of $\tilde{\bm\omega}$ and line 9 cannot lower it at $\varpi_t$ either, because $\tilde{\bm\omega}$ and $(d,\phi)$ are already optimal for the breaks of $\varpi_t$. Hence no single block update improves the criterion at the stopping point. \hfill$\blacksquare$

\subsection{Theorem~\ref{thm:detector} (Memory-Aware Detection)}
\textit{(i)} Let $\mathcal F_D=\sigma(\bar r_1,\dots,\bar r_D)$. Under $\Prob_\infty$, the Durbin--Levinson innovations are independent with $a_D\sim\mathcal N(0,\varsigma_D^2)$, and $a_D$ is independent of $\mathcal F_{D-1}$. Under a change of size $\delta$ on day $k$, the conditional law of $\bar r_i$ given $\mathcal F_{i-1}$ is normal with mean $\bar r_{i|i-1}+\delta\psi_{i,k}$ and variance $\varsigma_i^2$ for $i\ge k$. The likelihood ratio of $(\bar r_k,\dots,\bar r_D)$ given $\mathcal F_{k-1}$ is therefore $\ell_D(k,\delta)=\prod_{i=k}^{D}\exp[(\delta\psi_{i,k}a_i-\delta^2\psi_{i,k}^2/2)/\varsigma_i^2]$, and $\E_\infty[\ell_{D+1}(k,\delta)\mid\mathcal F_D]=\ell_D(k,\delta)$ for $k\le D$. A skipped day enters the statistic with a zero innovation, which replaces its factor by the deterministic $\exp[-\delta^2\psi_{i,k}^2/(2\varsigma_i^2)]\le1$, so with skipped days $\ell_D(k,\delta)$ is a nonnegative supermartingale. Integrating over $\delta\sim\mathcal N(0,\sigma_\delta^2)$ and using Fubini's theorem, $\Lambda_D(k)=\int\ell_D(k,\delta)\,d\mathcal N(0,\sigma_\delta^2)$ equals (\ref{eq:glr}), is a supermartingale in $D\ge k$, and satisfies $\E_\infty[\Lambda_k(k)\mid\mathcal F_{k-1}]\le1$. The Shiryaev--Roberts statistic~\cite{tartakovsky2014} $R_D=\sum_{k=1}^{D}\Lambda_D(k)$ thus satisfies $\E_\infty[R_{D+1}\mid\mathcal F_D]\le R_D+1$, so $R_D-D$ is a supermartingale with $R_0=0$. Let $T=\min\{D:R_D\ge A\}$. Optional stopping at the bounded time $T\wedge n$ gives $\E_\infty[T\wedge n]\ge\E_\infty[R_{T\wedge n}]\ge A\,\Prob_\infty(T\le n)$, because $R_D\ge0$ and $R_T\ge A$. Letting $n\to\infty$ gives $\E_\infty[T]\ge A$ when $\Prob_\infty(T<\infty)=1$, and $\E_\infty[T]=\infty$ otherwise. Restricting $k$ to $\mathcal K_D$ can only lower the maximum, and $\max_{k\in\mathcal K_D}\Lambda_D(k)\le R_D$, so the alarm day satisfies $N_A\ge T$ and $\E_\infty[N_A]\ge A$.

\textit{(ii)} For an autoregressive fractionally integrated moving average process ARFIMA$(0,d,0)$ with an infinite past, the innovation is $a_D=(1-B)^d\bar r_D$, the predictor is $\bar r_{D|D-1}=[1-(1-B)^d]\bar r_D$ and $\varsigma_D^2=\varsigma^2$. A unit step $\one\{D\ge\nu\}$ has the generating function $z^\nu/(1-z)$, so its image under $(1-B)^d$ has the generating function $z^\nu(1-z)^{d-1}$, whose coefficients are $\psi_{D-\nu}$ with $\psi_k=\binom{k-d}{k}=\Gamma(k+1-d)/[\Gamma(1-d)\Gamma(k+1)]$. Stirling's formula gives $\psi_k\sim k^{-d}/\Gamma(1-d)$. Under $\Prob_\nu$ the innovations are independent $\mathcal N(\delta\psi_{D-\nu},\varsigma^2)$, so the Kullback--Leibler divergence of $n$ post-change days is $I_n=\frac{\delta^2}{2\varsigma^2}\sum_{k<n}\psi_k^2$. For $d>0$, $\sum_{k<n}\psi_k^2\sim n^{1-2d}/[(1-2d)\Gamma(1-d)^2]$ by comparison with an integral, and for $d=0$, $\psi_k=1$ and $I_n=n\delta^2/(2\varsigma^2)$.

\textit{(iii)} Fix $\nu$, condition on $\mathcal F_{\nu-1}$ and on $\{N_A\ge\nu\}$, and take $D=\nu+n-1$ and $k=\nu$. Then $Q_D(\nu)=\sum_{j<n}\psi_j^2/\varsigma^2=2I_n/\delta^2$ and $Z_D(\nu)=\delta Q_D(\nu)+\sqrt{Q_D(\nu)}\,\mathcal G_n$ with $\mathcal G_n\sim\mathcal N(0,1)$ independent of $\mathcal F_{\nu-1}$. Substituting in (\ref{eq:glr}),
\begin{equation}
\log\Lambda_D(\nu)=\frac{\sigma_\delta^2Q_D(\nu)}{1+\sigma_\delta^2Q_D(\nu)}\Big(I_n+\delta\sqrt{Q_D(\nu)}\,\mathcal G_n+\frac{\mathcal G_n^2}{2}\Big)-\frac12\log\big(1+\sigma_\delta^2Q_D(\nu)\big)\ge I_n-\sqrt{2I_n}\,|\mathcal G_n|-O(\log I_n).\nonumber
\end{equation}
Since $I_n$ grows as $n^{1-2d}$ and $I_{n_A}\sim\log A$, for every $\epsilon>0$ there are $\eta_\epsilon>0$ and $A_\epsilon$ such that $I_n-\log A\ge\eta_\epsilon I_n$ for all $n\ge n_\epsilon=\lfloor(1+\epsilon)n_A\rfloor$ and $A\ge A_\epsilon$. If $k=\nu$ is a candidate on day $\nu+n-1$, then
\begin{equation}
\Prob_\nu(N_A-\nu+1>n\mid\mathcal F_{\nu-1},N_A\ge\nu)\le\Prob\big(\log\Lambda_{\nu+n-1}(\nu)<\log A\big)\le2\exp(-\eta_\epsilon^2I_n/16)\label{eq:tail}
\end{equation}
for $n\ge n_\epsilon$ and large $A$, uniformly in $\nu$ and $\mathcal F_{\nu-1}$, because the event requires $\sqrt{2I_n}\,|\mathcal G_n|>\eta_\epsilon I_n/2$ once the $O(\log I_n)$ term is below $\eta_\epsilon I_n/2$. If $K\ge cn_A$ with $c>1$, then $\nu$ is a candidate on day $\nu+n_\epsilon-1$ whenever $1+\epsilon\le c$, and the right side of \eqref{eq:tail} at $n=n_\epsilon$ tends to zero because $I_{n_\epsilon}\to\infty$. Since the delay is an integer, the events $\{N_A-\nu+1>(1+\epsilon)n_A\}$ and $\{N_A-\nu+1>n_\epsilon\}$ coincide. Larger values of $\epsilon$ follow by monotonicity, which proves the first claim. If every past day is a candidate, \eqref{eq:tail} holds for all $n\ge n_\epsilon$, so
\begin{equation}
\E_\nu[(N_A-\nu+1)^+\mid\mathcal F_{\nu-1}]\le n_\epsilon+\sum_{n\ge n_\epsilon}2\exp(-\eta_\epsilon^2I_n/16)=n_\epsilon+o(1),\nonumber
\end{equation}
because $I_n$ grows as a positive power of $n$ and $n_\epsilon\to\infty$. Since $\epsilon$ is arbitrary, the worst-case expected delay is at most $n_A(1+o(1))$. The asymptotic form of $n_A$ follows by inverting $I_n$ from (ii).

\textit{(iv)} The shift adds $\delta\psi_{i-\nu}$ to the one-step error of day $i$, so the excess squared error up to the alarm is $\delta^2\sum_{k=0}^{N_A-\nu}\psi_k^2=2\varsigma^2I_{N_A-\nu+1}$. Because $K\ge cn_A$, part (iii) gives $N_A-\nu+1\le(1+\epsilon)n_A$ with probability tending to one, and then $I_{N_A-\nu+1}\le(1+\epsilon)^{1-2d}\log A\,(1+o(1))$. For the lower bound, note that $\log\Lambda_D(k)\le\sigma_\delta^2Z_D(k)^2/[2(1+\sigma_\delta^2Q_D(k))]\le Z_D(k)^2/[2Q_D(k)]$. Under $\Prob_\nu$, $Z_D(k)/\sqrt{Q_D(k)}=\mu_k+\mathcal G_k$ with $\mathcal G_k\sim\mathcal N(0,1)$ and
\begin{equation}
\mu_k=\frac{\delta}{\varsigma}\frac{\sum_{i=\max(k,\nu)}^{D}\psi_{i-k}\psi_{i-\nu}}{\big(\sum_{i=k}^D\psi_{i-k}^2\big)^{1/2}},\qquad\text{so that}\qquad\mu_k^2\le\frac{\delta^2}{\varsigma^2}\sum_{j\le D-\nu}\psi_j^2=2I_{D-\nu+1}\nonumber
\end{equation}
by the Cauchy--Schwarz inequality, for candidates before and after the change alike. On days with $I_{D-\nu+1}\le(1-\epsilon)\log A$, an alarm requires $|\mu_k+\mathcal G_k|\ge\sqrt{2\log A}$ for some candidate $k$, hence $|\mathcal G_k|\ge(1-\sqrt{1-\epsilon})\sqrt{2\log A}$, an event of probability at most $2A^{-(1-\sqrt{1-\epsilon})^2}$. Because $K\le Cn_A$, there are at most $K(n_A+1)=O(n_A^2)$ pairs $(D,k)$ with $\nu\le D\le\nu+n_A$, a number that grows only polynomially in $\log A$, so the union bound vanishes as $A\to\infty$. Combining both bounds and letting $\epsilon\downarrow0$ gives $2\varsigma^2I_{N_A-\nu+1}=2\varsigma^2\log A\,(1+o_P(1))$. \hfill$\blacksquare$

\textit{Implementation.} The paper uses the stationary Durbin--Levinson predictor of order 60 days, a window of $K=60$ days and holiday skipping. Skipping and windowing are covered by part (i). With a finite order the innovations are exactly independent when the process is autoregressive of that order. For the fitted residual models of all test years, the partial autocorrelations of the daily mean residual beyond lag 60 do not exceed \PacfMax\ in absolute value, so the truncation changes the innovations only marginally.

\subsection{Theorem~\ref{thm:energy} (Energy Reserves Under Memory)}
\textit{(i)} The sum $\sum_te_t$ is $\mathcal N(0,V_n)$, so $\Prob(|\sum_te_t|\le E)=2\Phi(E/\sqrt{V_n})-1\ge1-\varepsilon$ if and only if $E\ge z\sqrt{V_n}$. With $E=z\sqrt{\Vind_n}$ the violation probability is $2\Phi(-z\sqrt{\Vind_n/V_n})$, which exceeds $2\Phi(-z)=\varepsilon$ if and only if $V_n>\Vind_n$, that is, $\sum_{s\neq t}\Sigma_{st}>0$.

\textit{(ii)} For a stationary sequence, $V_n=\sum_{|k|<n}(n-|k|)\gamma(k)$. Let $H\in(\frac12,1)$. By Lemma~\ref{lem:fou}, $\gamma(k)=C_\gamma k^{2H-2}(1+o(1))$ with $C_\gamma=\sigma^2H(2H-1)\kappa^{-2}$, so for every $\eta>0$ there is $k_0$ such that the terms with $k<k_0$ contribute $O(n)$ and the others lie within a factor $1\pm\eta$ of $2C_\gamma\sum_{k_0\le k<n}(n-k)k^{2H-2}$. A Riemann-sum argument gives $\sum_{k<n}(n-k)k^{2H-2}\sim n^{2H}\int_0^1(1-x)x^{2H-2}dx=n^{2H}/[(2H-1)2H]$, hence $V_n\sim\sigma^2\kappa^{-2}n^{2H}$, which dominates the $O(n)$ terms because $2H>1$. Since $\Vind_n=n\gamma(0)$, $\Vind_n/V_n\to0$ and the violation probability tends to $2\Phi(0)=1$. For $H=\frac12$, $\gamma(k)=\frac{\sigma^2}{2\kappa}e^{-\kappa|k|}$ and $V_n/n\to\sum_{k\in\mathbb Z}\gamma(k)=\frac{\sigma^2}{2\kappa}\frac{1+e^{-\kappa}}{1-e^{-\kappa}}=\frac{\sigma^2}{2\kappa}\coth\frac\kappa2$.

\textit{(iii)} Under the affine policy, the thermal imbalance $(1-b_h)e_h$ is $\mathcal N(0,(1-b_h)^2\hat\sigma_h^2)$, the storage imbalance $b_he_h$ is $\mathcal N(0,b_h^2\hat\sigma_h^2)$, and the cumulative storage imbalance after $n$ hours, $\sum_{h\le n}b_he_h=(\bm b\odot\one_{1:n})^\T\bm e$, is $\mathcal N(0,\|\bm U(\bm b\odot\one_{1:n})\|_2^2)$ because $\hat{\bm\Sigma}_D=\bm U^\T\bm U$. By (i), each two-sided chance constraint of level $\varepsilon_P$ or $\varepsilon_E$ is equivalent to the corresponding constraint of (\ref{eq:socp}). The functions $(1-b_h)\hat\sigma_h$ and $b_h\hat\sigma_h$ are affine and $\|\bm U(\bm b\odot\one_{1:n})\|_2$ is convex in $\bm b$, so the feasible set is second-order cone representable and (\ref{eq:socp}) is a second-order cone program. \hfill$\blacksquare$

\section{Numerical Checks}\label{sec:checks}
Table~\ref{tab:checks} compares each closed form used in the proofs with an independent computation by numerical integration, power-series inversion or direct summation. The largest errors are taken over $\alpha\in\{0.3,0.55,0.8\}$ and $x$ from 0.01 to 100 for Proposition~\ref{prop:colecole}, over $\alpha\in\{0.3,0.57,0.8\}$, $\bar\tau\in\{2,7.8,30\}$~h and seven lags up to 2000 for the mixture of Theorem~\ref{thm:fes}(i), and over $\kappa\in\{0.05,0.3,1\}$, five Hurst exponents from 0.15 to 0.85 and six lags up to 40 for Lemma~\ref{lem:fou}. The kernel checks of Theorem~\ref{thm:fes}(i) and (iii) use the heating kernel estimated for ISO New England (ISO-NE), with $\alpha=0.57$ and $\bar\tau=7.8$~h. 

\begin{table}[!ht]
\caption{Numerical Checks of the Closed Forms Used in the Proofs}\label{tab:checks}
\centering\footnotesize
\begin{tabular}{@{}lp{0.6\textwidth}l@{}}
\toprule
Result & Comparison & Value\\
\midrule
Proposition 1(a) & Stieltjes transform of $f_\alpha$ and $1/(1+x^\alpha)$, largest error & $3.2\times10^{-11}$ \\
Proposition 1(a) & Characteristic function of $V_i$ and $\sinh(\pi\omega)/[\alpha\sinh(\pi\omega/\alpha)]$, largest error & $6.5\times10^{-14}$ \\
Proposition 1(c) & $\Var(\ln\tau_i)$ and $\pi^2(\alpha^{-2}-1)/3$, largest error & $1.4\times10^{-14}$ \\
Theorem 1(i) & $w_n$ and $\E[\lambda_V(1-\lambda_V)^n]$, largest relative error & $4.9\times10^{-8}$ \\
Theorem 1(i) & Smallest $(-\Delta)^kw_n$ for $k=1,2,3$ and $n<3000$, ISO-NE heating kernel & $1.7\times10^{-15}$ \\
Theorem 1(ii) & Smallest weight for $\alpha=1.2$, 1.5 and 1.8 & $-$0.0020, $-$0.0146, $-$0.0486 \\
Theorem 1(iii) & $\bar w_n$ over $c\,n^{-\alpha}/\Gamma(1-\alpha)$ at $n=10^3$, $10^4$, $10^5$, same kernel & 1.014, 1.004, 1.001 \\
Theorem 1(iii) & $w_n$ over $\alpha c\,n^{-1-\alpha}/\Gamma(1-\alpha)$ at the same $n$ & 1.027, 1.008, 1.002 \\
Lemma 1 & Closed form and spectral integral of $\gamma(s)$, largest relative error & $1.0\times10^{-10}$ \\
Theorem 2(ii) & Signature of the order-400 predictor and $\psi_k$, $k\le50$, $d=0.3$, largest error & $4.0\times10^{-3}$ \\
Theorem 3(ii) & $V_n$ over $\sigma^2\kappa^{-2}n^{2H}$ at $n=2000$, 8000, 32000, $H=0.75$, $\kappa=0.2$ & 0.9999, 1.0000, 1.0000 \\
 
\bottomrule
\end{tabular}
\end{table}

\section{Settings}\label{sec:settings}
\subsection{Proposed Framework}
\textit{Kernel.} The weights $w_n$ of (\ref{eq:fes}) are computed by power-series inversion for $n<4096$, and the tail mass $\bar w_{4095}$ is assigned to the input 4096 hours earlier, so that the filter keeps unit gain exactly. Filtering uses the fast Fourier transform and runs on the full temperature history, so the memory of the kernel is never cut at the start of a training window.

\textit{Backbone.} The base temperatures are $T^{\HD}=15.5\,^\circ$C and $T^{\CD}=18.5\,^\circ$C. The calendar block has 168 hour-of-week indicators, 24 holiday-by-hour indicators, $\sin$ and $\cos$ of $2\pi k\,\mathrm{doy}/365.25$ for $k=1,2,3$ interacted with the 24 hours, where doy is the day of year, and a linear trend in years. The profile search runs two passes of a coordinate grid over $\alpha\in\{0.40,0.55,0.70,0.85,1\}$ and $\bar\tau\in\{3,6,12,24,48\}$~h for heating and cooling, followed by at most 150 Nelder--Mead iterations~\cite{nelder1965} in logit-log coordinates. The profile is computed on all years up to $Y-2$ and kept for the calibration and final models of test year $Y$.

\textit{Boosting.} The boosted correction uses LightGBM with 1200 trees, learning rate 0.03, 63 leaves, at least 50 samples per leaf, row and column subsampling of 0.8, $L_2$ penalty 1 and seed 0.

\textit{Breaks.} In the daily model (\ref{eq:daily}), $\tilde{\bm x}_D$ contains six weekday indicators, a holiday indicator, two annual harmonic pairs and a linear trend. The annual estimation uses $\ell=180$ days and $m_{\max}=6$, and the full-sample analysis uses $\ell=180$ days for ISO-NE and 90 days for PJM Interconnection. The grids are $\mathcal G_d=\{0,0.05,\dots,0.45\}$ and $\mathcal G_\phi=\{0,0.1,\dots,0.9\}$, and the fractional filter is truncated at 400 days. The inner loop of Algorithm~\ref{alg:joint} runs at most 4 times, and the outer loop at most 6 times in the annual estimation and 8 times in the full-sample analysis. When the inner loop ends with unchanged breaks and line 9 returns the current memory parameters, the algorithm stops at once, since the next iteration would repeat the tuple. Confidence intervals of break dates follow Bai~\cite{bai1997} with a Bartlett long-run variance. The training window is the current regime extended backwards until it covers at least 365 days.

\textit{Residual model.} The hourly scales $\vartheta_h$ come from the calibration year. Covariance matching compares the $24\times24$ matrices of model and sample second moments at day lags 0 to 14, with weight $\frac12$ on the symmetric lag-0 matrix, whose entries appear twice. Each component is parametrized by its variance $v\in[10^{-4},2]$, its rate $\kappa\in[10^{-3},5]$ per step (an hour for $X$ and a day for $P$ and $M$) and $H\in[0.05,0.95]$, all mapped from $\mathbb R$ by logistic transformations, and $\sigma^2=v\kappa^{2H}/[H\Gamma(2H)]$. The best of three Nelder--Mead starts is kept. The conditioning set has 346 hours.

\textit{Detector.} The detector monitors the daily mean residuals with the autocovariance implied by the fitted residual model, a stationary Durbin--Levinson predictor of order 60, a window of $K=60$ days, which covers shifts of two innovation standard deviations or more by Example~\ref{ex:delay}, at least 3 post-change days, a prior standard deviation $\sigma_\delta$ equal to 2\% of the mean training load and $A=1826$ days. Holidays and the days around them are skipped, and the last 60 days of the calibration year serve as a warm-up.

\textit{Reserves.} The risk levels are $\varepsilon_P=\varepsilon_E=0.05$, the round-trip efficiency is $\eta=0.79$, $\pi^S=1$~\$/MW-h and the penalty is \$1000/MWh. Program (\ref{eq:socp}) is solved by CLARABEL through CVXPY as a parametrized program that is compiled once. Benchmarks without a covariance enter the program with the diagonal Gaussian law whose hourly standard deviations match their 5\% and 95\% quantiles. A power shortfall is the excess of $|(1-b_h)e_h|$ over $R_h$ plus that of $|b_he_h|$ over $S_h$, and an energy shortfall accrues whenever the absolute cumulative storage imbalance grows beyond $E$, by the amount of that growth. The daily energy violation probability under each forecaster's own law is estimated with 4000 standard normal draws shared by all days and methods.

\subsection{Benchmarks}
All benchmarks use the information set of the paper. The vanilla multiple linear regression (MLR)~\cite{hong2010} has trend, month, weekday-by-hour and holiday-by-hour terms and hour- and month-specific cubic polynomials of temperature. The recency MLR~\cite{wang2016} adds hour-specific cubic polynomials of the temperatures lagged by 1, 2, 3, 4, 6, 8 and 12~h and of the 24-h moving averages of the current and previous day. The adaptive generalized additive model (GAM) has hour-by-day-type indicators (weekday, weekend and holiday), cubic B-splines of temperature, of exponentially smoothed temperature (factor 0.05) and of temperature by three-hour block, and two annual harmonic pairs. It is estimated offline by least squares and then tracked by a Kalman filter with random-walk state noise $q\bm I$, where $q\in\{0,10^{-7},10^{-6},10^{-5}\}$ is selected on year $Y-1$, and it is updated every hour until the issue time. For adaptive probabilistic load forecasting (APLF)~\cite{alvarez2021}, we reimplemented the published method with both forgetting factors 0.2 and 0.7 and 48 calendar types, and we predict recursively to the end of day $D$. The gradient-boosted decision trees (GBDT) use LightGBM with 1500 trees, learning rate 0.03 and 63 leaves on the hour, weekday, day of year, holiday indicator, temperature, dew point, temperatures lagged by 1, 2, 3, 6, 12, 24 and 48~h and their moving averages over 6, 24, 72 and 168~h. GBDT with load lags adds the latest same-hour load available at the issue time (one or two days earlier), the same hour one week earlier, and the last load and the 24-h mean load at the issue time.

The neural networks take the 168 hourly loads and temperatures up to the issue time with hour encodings, the 24 temperatures and dew points of day $D$, and weekday, holiday and seasonal terms, and they output the 24 loads of day $D$. The multilayer perceptron (MLP) has two hidden layers of 256 units with dropout 0.1 and 99 quantile outputs trained with the pinball loss. The long short-term memory (LSTM) network has two layers of 64 units with dropout 0.1 over the seven days of the past week, each step carrying the 24 hourly loads, temperatures and hour encodings of one day, followed by a dense head of 128 units that also receives the exogenous inputs. The patch transformer has seven daily patches and a target token and two encoder layers of width 64 with four heads and a feed-forward width of 128. The Kolmogorov--Arnold network (KAN) has a linear embedding of width 48 followed by two Kolmogorov--Arnold layers ($48\to32\to24$) with cubic B-splines on six grid intervals and a base branch with the sigmoid linear unit. All networks are trained by Adam with learning rate $10^{-3}$, weight decay $10^{-5}$ and batches of 64 days for at most 80 epochs, with early stopping (patience 12) on the last 10\% of the training period.

The intervals of the LSTM network, the transformer and the KAN are hour-specific empirical quantiles of the errors on that held-out segment. The MLR, GBDT and APLF benchmarks receive hour-specific empirical quantiles of the errors of year $Y-1$ predicted by a model trained up to $Y-2$, and the adaptive GAM uses its Kalman predictive variance. APLF receives these intervals because its recursive Gaussian variance, designed for short horizons, collapses at lead times of 15 to 38~h.

For adaptive multitask learning~\cite{zaballa2026} we reimplemented the published method with both forgetting factors equal to 0.9, 48 calendar types and a correlation threshold of 0.1, and warmed it up on 2015.

\section{Data Processing}\label{sec:data}
\textit{ISO-NE.} The yearly standard market design workbooks of 2003 to 2017, as distributed with the data package of the Global Energy Forecasting Competition 2017~\cite{hong2019}, provide the hourly system and zonal demand, dry-bulb and dew-point temperatures, day-ahead and real-time locational marginal prices and regulation clearing prices. Temperatures are converted from Fahrenheit to Celsius. The workbooks report 24 hours per day. The zero demand that the early workbooks report in the spring daylight-saving hour is replaced by linear interpolation, as are isolated missing values. Holidays follow the competition's holiday list, which starts in 2005, completed for 2003 and 2004 by the United States federal holiday rules. The system series starts on 1 March 2003, when the standard market design began.

\textit{PJM.} The data set of Ruan \emph{et al.}~\cite{ruan2020} provides the hourly metered load of the PJM footprint, day-ahead prices, and temperature and dew point from three airport stations (Philadelphia, Columbus and Richmond) from January 2017 to November 2021. Gaps of up to 48 hours are interpolated, and longer gaps would be filled with the values of the previous week, although the released series needed no such filling. Holidays follow the United States federal rules.

\textit{Zones.} The eight zones of the competition use their own load and temperature columns of the same workbooks. The zone order in Table~\ref{tab:zones} is alphabetical and matches the entity order of Zaballa \emph{et al.}~\cite{zaballa2026}, which we confirmed against the first hourly values of the data file released with their method.

\section{Additional Results}\label{sec:more}
Table~\ref{tab:peryear} reports the mean absolute percentage error (MAPE) of every method in every test year, over the same days as Table~\ref{tab:main}. The proposed method has the lowest MAPE in all nine ISO-NE years and in 2020 and 2021 in PJM, and the LSTM network has the lowest in PJM in 2019. Table~\ref{tab:alarms} lists all alarms of the online detector, with the documented events that lie near the estimated start of each shift, such as the restrictions of the coronavirus disease 2019 (COVID-19) pandemic.

\begin{table}[!ht]
\caption{MAPE (\%) per Test Year}\label{tab:peryear}
\centering\footnotesize
\setlength{\tabcolsep}{3.6pt}
\begin{tabular}{@{}lccccccccccc@{}}
\toprule
 & Vanilla & Recency & Adaptive & APLF & GBDT & GBDT & MLP & LSTM & Patch & KAN & Proposed\\
Year & MLR & MLR & GAM & & & with lags & & & transformer & & \\
\midrule
ISO-NE 2008 & 3.44 & 2.75 & 2.61 & 4.22 & 1.80 & 1.55 & 1.65 & 1.57 & 1.59 & 1.85 & \textbf{1.47} \\
ISO-NE 2009 & 4.69 & 4.17 & 2.63 & 4.05 & 3.56 & 2.14 & 1.97 & 2.08 & 2.09 & 2.25 & \textbf{1.50} \\
ISO-NE 2010 & 2.94 & 2.34 & 2.83 & 4.73 & 2.60 & 1.84 & 2.09 & 1.84 & 1.98 & 2.17 & \textbf{1.59} \\
ISO-NE 2011 & 3.14 & 2.56 & 3.00 & 5.01 & 3.05 & 2.06 & 2.12 & 1.98 & 2.26 & 2.17 & \textbf{1.83} \\
ISO-NE 2012 & 3.00 & 2.45 & 2.78 & 4.62 & 3.11 & 1.77 & 1.78 & 1.62 & 1.81 & 2.09 & \textbf{1.51} \\
ISO-NE 2013 & 3.20 & 2.53 & 2.65 & 4.67 & 2.69 & 1.73 & 1.66 & 1.66 & 1.79 & 1.85 & \textbf{1.52} \\
ISO-NE 2014 & 3.36 & 2.78 & 2.77 & 4.81 & 3.04 & 1.82 & 1.52 & 1.55 & 1.67 & 1.85 & \textbf{1.37} \\
ISO-NE 2015 & 3.88 & 3.10 & 2.87 & 5.10 & 4.27 & 2.15 & 1.99 & 1.97 & 1.90 & 2.06 & \textbf{1.51} \\
ISO-NE 2016 & 4.04 & 3.37 & 2.96 & 5.31 & 5.86 & 2.39 & 2.11 & 2.36 & 2.22 & 2.49 & \textbf{1.72} \\
PJM 2019 & 4.95 & 3.75 & 3.13 & 5.40 & 2.51 & 2.34 & 2.61 & \textbf{2.18} & 2.57 & 2.85 & 2.26 \\
PJM 2020 & 4.49 & 3.77 & 3.48 & 5.33 & 3.85 & 2.66 & 2.58 & 2.85 & 2.94 & 3.61 & \textbf{2.15} \\
PJM 2021 & 3.85 & 3.37 & 3.01 & 5.63 & 2.32 & 2.11 & 2.09 & 2.28 & 2.34 & 2.29 & \textbf{1.97} \\
 
\bottomrule
\multicolumn{12}{@{}l}{\scriptsize Bold marks the lowest MAPE of each year.}
\end{tabular}
\end{table}

\begin{table}[!ht]
\caption{Alarms of the Online Detector}\label{tab:alarms}
\centering\footnotesize
\begin{tabular}{@{}llccl@{}}
\toprule
System & Alarm day & Estimated start & $\hat\delta$ (GW) & Documented event near the estimated start\\
\midrule
ISO-NE & 2008-12-14 & 2008-12-12 & $-$1.05 & ice storm (11--12 Dec. 2008) \\
ISO-NE & 2008-12-21 & 2008-12-15 & 0.66 & restoration after the ice storm \\
ISO-NE & 2010-02-04 & 2010-01-03 & 0.43 & -- \\
ISO-NE & 2010-06-04 & 2010-06-02 & 0.62 & -- \\
ISO-NE & 2010-06-09 & 2010-06-07 & $-$0.76 & -- \\
ISO-NE & 2011-08-29 & 2011-08-27 & $-$1.10 & tropical storm Irene (28 Aug. 2011) \\
ISO-NE & 2011-09-08 & 2011-09-01 & 0.77 & restoration after Irene \\
ISO-NE & 2011-10-30 & 2011-10-28 & $-$1.13 & snowstorm (29--30 Oct. 2011) \\
ISO-NE & 2011-11-05 & 2011-11-03 & 0.96 & restoration after the snowstorm \\
ISO-NE & 2015-01-29 & 2015-01-27 & $-$0.67 & blizzard (26--28 Jan. 2015) \\
ISO-NE & 2015-02-01 & 2015-01-30 & 0.87 & recovery after the blizzard \\
PJM & 2019-04-21 & 2019-03-14 & $-$3.13 & -- \\
PJM & 2020-04-05 & 2020-03-09 & $-$3.89 & COVID-19 restrictions \\
PJM & 2020-06-22 & 2020-06-02 & 4.42 & reopening after COVID-19 restrictions \\
PJM & 2021-07-17 & 2021-07-08 & $-$5.03 & -- \\
PJM & 2021-07-22 & 2021-07-18 & 6.56 & -- \\
 
\bottomrule
\multicolumn{5}{@{}l}{\scriptsize A dash marks an alarm without a documented event within a few days of its estimated start.}
\end{tabular}
\end{table}

\begingroup
\papersizes
\ifdefined\AddToHookNext\AddToHookNext{shipout/after}{\gdef\leftmark{}\gdef\rightmark{}}\fi%
\begin{multicols}{2}

\end{multicols}
\endgroup

\end{document}